\documentclass[%
 reprint,
superscriptaddress,
 amsmath,amssymb,
 aps,
prx,
]{revtex4-2}

\usepackage{amsmath,amssymb}
\usepackage{graphicx,float}
\usepackage{times,txfonts}
\usepackage{color}
\usepackage{multirow}
\usepackage{bbold}
\usepackage{color}
\usepackage{soul}
\usepackage{subfigure}

\usepackage{hyperref}
\usepackage{times,txfonts}
\usepackage{float}
\usepackage{natbib}
\usepackage{nicefrac}
\usepackage{blindtext}
\usepackage[export]{adjustbox}
\usepackage{mathrsfs}
\usepackage{textcomp}
\usepackage{blkarray}
\usepackage{multirow}

\renewcommand{\epsilon}{\varepsilon}

\def\VR{\kern-\arraycolsep\strut\vrule &\kern-\arraycolsep}
\def\vr{\kern-\arraycolsep & \kern-\arraycolsep}
\usepackage{sistyle}
\usepackage{soul}
\definecolor{lightblue}{RGB}{185,210,248}
\sethlcolor{lightblue}

\begin{document}

\title{Satellite Relayed Global Quantum Communication without Quantum Memory}

\author{Sumit Goswami }
\thanks{sumitsgoswami@gmail.com}
\affiliation{Institute  for  Quantum  Science  and  Technology  and  Department  of  Physics  and  Astronomy, University  of  Calgary, Calgary  T2N  1N4, Alberta, Canada}

\author{Sayandip Dhara}
\thanks{sayaniiser1@gmail.com}
\affiliation{Department of Physics, University of Central Florida, Orlando, Florida 32816, USA}

\begin{abstract}

Photon loss is the fundamental issue toward the development of quantum networks. To circumvent loss quantum repeaters were proposed, which need high-performance quantum memories. We present an alternative proposal to mitigate photon loss even at large distances and hence to create a global scale quantum communication architecture without quantum repeaters. In this proposal, photons are sent directly through space, using a chain of co-moving low-earth orbit satellites. This satellite chain would bend the photons to move along the earth's curvature and control photon loss due to diffraction by effectively behaving like a set of lenses on an optical table. Numerical modeling of photon propagation through these 'satellite lenses' shows that diffraction loss in entanglement distribution can be almost eliminated even at global distances of 20,000 km while considering beam truncation at each satellite and the effect of different errors (e.g. 'satellite lens' focal length fluctuation). In absence of diffraction loss, the effect of other losses (especially reflection loss) becomes important and they are investigated in detail. The total loss is estimated to be less than 30 dB at 20,000 km if other losses are constrained to 2$\%$ at each satellite, with 120 km satellite separation and 60 cm diameter satellite telescopes eliminating diffraction loss. Such low loss satellite-based optical-relay protocol would enable robust, multi-mode global quantum communication and wouldn't require either quantum memories or repeater protocol. The protocol can also be the least lossy in almost all distance ranges available (200 - 20,000 km). Recent advances in space technologies may soon enable affordable launch facility for such a satellite-relay network. We further introduce the ‘qubit transmission’ protocol which has a plethora of advantages with both the photon source and the detector remaining on the ground. A specific lens setup was designed for the 'qubit transmission' protocol which performed well in simulation that included atmospheric turbulence in the satellite uplink.

\end{abstract}

\maketitle
%%%%%%%%%%%

\section{Introduction}\label{intro}

A global quantum network would enable us to transfer quantum information between any two place on earth. The development of a worldwide quantum network would be immensely important for emerging quantum technologies as well as be a boon to fundamental research \cite{yin_satellite-based_2017,xu_satellite_2019}. A functioning quantum network would facilitate secure communication through global quantum key distribution (QKD) \cite{bennett_quantum_1992,wehner_quantum_2018,yin_entanglement-based_2020}, distributed quantum computing, and entanglement-based precision quantum sensing useful in atomic clocks, very long baseline telescopes, etc \cite{gottesman_longer-baseline_2012}.  The fundamental issue in constructing a quantum network is photon loss as light can not be amplified (as done in the classical network) in quantum communication due to no-cloning theorem \cite{wootters_single_1982}. Following classical networks, in a quantum network, the default choice for qubit transmission has been optical fibers which can have an extremely low loss (around 0.15 dB/kilometer) \cite{simon_towards_2017}. However, even this small loss is especially difficult for the single photons needed to be transmitted in a quantum network because absorption loss scales exponentially with distance. Even using the best quantum sources available, it would require more time than the age of the universe to transmit a single photon through 2000 kilometers. This led to different strategies for building quantum network at different distances. Until 500 km qubit transmission through optical fiber seemed the best. For further distances, two separate strategies are being explored. In one strategy, photons are still transmitted through optical fibers while clever quantum repeater protocols were designed to get around the problem of photon loss using quantum memories \cite{Briegel98,Sangouard}. The other strategy - gaining prominence over the last decade - is sending photons to or from orbiting satellites \cite{yin_satellite-based_2017,liao_satellite--ground_2017, ren_ground--satellite_2017, liao_satellite-relayed_2018,yin_entanglement-based_2020, chen_integrated_2021}. 

The main roadblock for implementing quantum repeaters has been the requirement of high-performance quantum memories, although many other problems also persist \cite{simon_towards_2017}. There has been now almost two decades of research on quantum memories that has yielded separately high efficiency, high storage time , and moderate multimode capacity memories \cite{wang_efficient_2019,yang_efficient_2016,pu_experimental_2017}. However, for a practical quantum repeater, high capabilities in all these memory characteristics are needed together, which still seems rather difficult. Due to this, functional quantum repeaters are currently limited to distances even below 100 km \cite{yu_entanglement_2020, pompili_realization_2021, valivarthi_teleportation_2020}. Quantum memories also do not have simple and robust designs yet that can be easily deployed commercially, mostly requiring sophisticated setups and cryogenic operational temperatures. On the other hand, photon transmission through quantum satellites has garnered some spectacular success over the recent years \cite{yin_satellite-based_2017,liao_satellite--ground_2017, ren_ground--satellite_2017, liao_satellite-relayed_2018,yin_entanglement-based_2020, chen_integrated_2021}, with successful demonstration of entanglement distribution up to 1200 km on earth using Micius satellite in Low earth orbit (LEO) \cite{yin_satellite-based_2017}. It was achieved since photon transmission through space faces diffraction loss, which scales quadratically with distance as compared to the exponential scaling of the absorption loss in fiber.

However, neither fiber-based repeaters nor LEO satellites can transfer quantum information very far beyond 2000 km \cite{simon_towards_2017}. At large distances, quantum repeaters become very complicated requiring many links, although some proposals exist \cite{zwerger_long-range_2018}. LEO satellites, being on low elevation (200-2000 km) compared to earth radius (around 6400 km), faces the curvature of earth quickly. To reach further distances (2000-20,000 km), either higher orbit satellites (e.g., geostationary satellites at 36,000 km elevation) or schemes combining quantum memory with satellite are proposed \cite{Mustafa}. But both these proposals have small transmission rates due to either photon loss in diffraction or long storage times in the memory\cite{boone_entanglement_2015}.

In our proposal, a chain of closely spaced co-moving LEO satellites is used to transmit the photonic qubits directly through space without using quantum memories or repeater protocols. Thus, the satellite chain would create an All-Satellite Quantum Network, called ASQN henceforth. In ASQN, photons are simply reflected from one satellite to another using satellite telescopes. The LEO satellite chain would transmit photons by bending light along the curvature of the earth and thereby mitigating the huge diffraction losses faced in transmission from higher orbit satellites. Moreover, diffraction is only beam divergence which, as we know, can be controlled using optical elements. In that effect, telescope mirrors in satellites can be effectively used as lenses to focus incoming light beams. The satellites are chosen to be co-moving in the same orbit (i.e. stationary relative to each other) and hence the chain of satellites would effectively behave as a set of lenses on an optical table that converges the light periodically to completely contain beam diffraction. As each satellite effectively behaves as a lens, we would call them 'satellite lenses'. For small enough satellite separation distance ($\sim$ 120 km) and large enough telescope diameter or ‘satellite lens’ size  ($\sim$ 60 cm), diffraction loss can be eliminated. Analytical treatment of light propagation helps us determine the focal lengths of the 'satellite lenses'. Numerical modelling in Section \ref{result_simulation} simulates diffraction loss by considering beam truncation due to finite size of the telescope mirrors. It shows that photons can be transmitted to even large global distances (upto 20,000 km) with almost no diffraction loss. Hence, ASQN is especially important for long-range (5,000-20,000 km) quantum communication.

Due to the complete absence of diffraction loss, other forms of losses become dominant. The two principal forms of losses to consider are light absorption at each satellite and effect of other factors causing additional diffraction loss. Absorption loss - just like in optical fiber and in air transmission - increases exponentially in the satellite chain too, with increasing number of satellites \cite{ursin_entanglement-based_2007}. Hence, light absorption at each satellite must be kept to a minimum to stop the total absorption loss from growing dangerously high. To achieve such low loss only optical elements that can be used in all satellites are reflectors. Other optical elements are generally very lossy (e.g., loss due to the front surface reflection and absorption in a glass lens). So, they cannot be used in the light path in all satellites, although a few satellites - like the ground pointing satellites - may use them if needed. In all the satellites, only reflecting telescopes are used to just collect and transmit the light while providing focusing to the beam as needed. Even reflection loss itself can quickly become extreme with a large number of satellites. Hence, reflection loss is one of the most important factors in ASQN. We considered reflection loss from from metal mirrors and also Bragg mirrors, which are almost ideal reflectors with near unity reflectivity. The different challenges and advantages associated with both these options were considered in Section \ref{mirrors}.

There are multiple other factors that may contribute to loss by affecting beam diffraction. These factor were investigated in quite some detail and whenever applicable possible solutions were suggested to limit their adverse effects. For example, to limit beam aberration we have suggested different telescope setups and using vortex beam transmission through on-axis telescope setup (as discussed in Section \ref{telescope_setups}). Focal length adjustment was discussed in details too. Another source of loss is satellite chain setup error which consists of satellite tracking error, satellite position error, focusing error etc. The effect of some of these different sets of errors on diffraction loss is considered in Section \ref{error_appen}. The most important factor is the tracking error which can be very challenging to control over a satellite chain containing hundreds of satellites if all satellites need to be dynamically tracked. However, in ASQN all the satellites in the chain move in almost identical orbits (co-moving) and hence they will be stationary with respect to each other. So, the satellites don't need to be tracked at all or tracked minimally, except at the ground links. Instead, the ‘stationary satellite lenses’ only need to be aligned which would cause much less error. 

The total loss due to absorption loss, setup error, and other losses (like atmospheric loss in ground link) together are considered to the best possible extent. The total loss progressively increases with distance predominantly due to reflection loss at each satellite. Considering 2 $\%$ reflection loss at each satellite, the total loss in entanglement distribution turned out to be only around 27 dB even at 20,000 km, while with conservative parameters the loss is still 50 dB (as shown in Fig. \ref{Result_Fig3} in Section \ref{result_simulation}). It needs to be noted reflection loss does not have a fundamental lower bound. If large Bragg mirrors can be used in the telescopes, reflection loss would completely vanish due to high reflectivity of Bragg mirrors ( $>$ 99.99 $\%$). In that case, ASQN would have ultra-low total loss - only around 10 dB of overhead losses - even at global distances.

However, a large scale satellite-relay like ASQN would have many facades. Everything can not be analyzed in this work, and to ultimately decide on its feasibility further research including both theoretical analysis (e.g. to ascertain precise effects of aberration) and possible tabletop experiments would be required (as discussed in \ref{tabletop}).

In the regular entanglement distribution protocol, two photons belonging to the entangled pair are transmitted along two directions using reflector satellites. As ASQN consists of reflector satellites, we also propose the first qubit transmission protocol through space. In qubit transmission (Fig. \ref{scheme}(a)), photons are sent from a source kept in a ground station to the detector in another ground station. Hence, instead of having either source or detector on the satellite, they both stay in the ground station, which produces a slew of advantages operationally (discussed in detail in Section \ref{scheme}). However, qubit transmission has its challenge with turbulence in the uplink which widens the beam and can decrease transmission rate significantly \cite{fante_electromagnetic_1975,fante_electromagnetic_1980,bonato_feasibility_2009,pugh_adaptive_2020}. The proposed lens setup deals with this issue and despite turbulence numerical simulation in Section \ref{result_turb} shows a reasonably high rate of transmission even at global distances.

Both entanglement distribution and qubit transmission protocols can create a quantum network to execute QKD in global distances. ASQN loss is below 30 dB at 20,000 km and optical fiber loss is at 0.15 dB/kilometer. Also, loss in ASQN is smaller at shorter distances. Hence, ASQN will perform better than even direct optical fiber transmission for distances around 200 km  - around the distance ASQN itself starts to become meaningful with two links of around 100 km. This means that ASQN would not only be important in large distances. It can be the preferred quantum communication protocol with least loss over almost the whole range of distances (200- 20,000 km) it will function. A single quantum communication protocol with such low loss over almost the complete distance range in earth would be unprecedented.

ASQN will not require either quantum memories or repeater protocol. As stated earlier, building high performance memories has been a major roadblock for memory-based quantum network protocols which can be avoided in ASQN. Also in absence of memories, low temperature operations will not be necessary in the satellites and large multimode capacity would be much easier to build. Memories will however be important in building more sophisticated operations like a quantum internet \cite{wehner_quantum_2018} (See Section \ref{QI_appen}) in ASQN. Quantum memories and fiber based quantum repeater protocols would be important to boost capacity too. This would be especially important at short distances (below 2000 km, say) where a web of ground based fiber network can achieve much higher capacity transmission than a satellite-chain network.

Recently free space diffraction-controlled quantum communication schemes using drones, similar to the this proposal, were experimentally demonstrated \cite{liu_optical-relayed_2021,conrad_drone-based_2021}. This shows the experimental feasibility of such a scheme although the experiments only show transmission through a small distance (up to 1 km \cite{liu_optical-relayed_2021}) in the air using drones with a very different relay technology using fiber mode matching and polarization correctors, which are both quite lossy. The possibility of using satellites was mentioned but rejected in \cite{liu_optical-relayed_2021},  by only considering long-distance links (around 1000 km) which would then need large telescopes. A long chain of satellites was not considered possibly because of the large loss suffered by each satellite in their scheme due to mode matching, polarization correcting instruments, and tracking error. In ASQN, all these sources of loss are accounted for. Fiber mode matching is not needed as we are using mirror reflections, temporal or frequency qubits are used instead of polarization qubits, and co-moving (i.e. relatively stationary) satellite chain makes dynamic tracking irrelevant. These reasons made a long chain of closely separated satellites possible which can also be at a low elevation - consequently reducing diffraction loss in ground transmission - as each satellite does not need to cover a large area on the ground due to the small separation between them.

Constellation of quantum satellites has already been planned and is being extensively studied for providing long distance trusted node QKD services \cite{brito_satellite-based_2021}. However, launching all these satellites equipped with moderately sized telescopes would be another technical challenge of ASQN. To create a global quantum link spanning around 20,000 km hundreds of satellites are needed for a single chain (e.g., transmission at 800 nm wavelength using 60 cm diameter telescopes would require $\sim$ 120 km satellite separation i.e., $\sim$ 160 satellite). Multitudes more are needed for a constellation of satellites always providing quantum connection between any two points of earth. However, with recent rapid advances in space technology this may be possible in the near future. The development of reusable rockets technology is dramatically increasing accessibility to space. Banking on this, multiple classical internet satellite networks have been announced recently \cite{noauthor_euroconsult_nodate}. Some of these are already operational with thousands of satellites already launched \cite{noauthor_oneweb_2017,noauthor_amazon_2019,noauthor_arianespace_2021,noauthor_euroconsult_nodate,noauthor_spacex_2021} and tens of thousands more planned to be launched \cite{noauthor_spacex_2019}. With efforts underway to build very large fully reusable spaceships suitable for interplanetary travel \cite{noauthor_spacex_2021-1}, access to space can soon increase even more dramatically as cost decreases.

The paper is structured as the following. After introduction in Section \ref{intro} we describe ASQN in detail in Section \ref{scheme}. We analyzed ASQN both theoretically and numerically in Section \ref{Results} and results are presented for both entanglement distribution and qubit transmission protocols. Further, the effect of uplink turbulence on the qubit transmission protocol was numerically modeled. Influence of different elements on  the above results were discussed in Section \ref{Discussion}, possible tabletop experiments to verify different aspects of ASQN are described in Section \ref{tabletop} and conclusions were depicted in Section \ref{concl} . Finally in Appendix (Section \ref{appendix}) we discuss the details of implementing full quantum internet capabilities using our proposal and analyze the effect of satellite chain setup errors.

\section{Scheme}\label{scheme}

\begin{figure}[htbp]
    \centering
    \begin{subfigure}{}
      \centering
      \includegraphics[width=.9\linewidth]{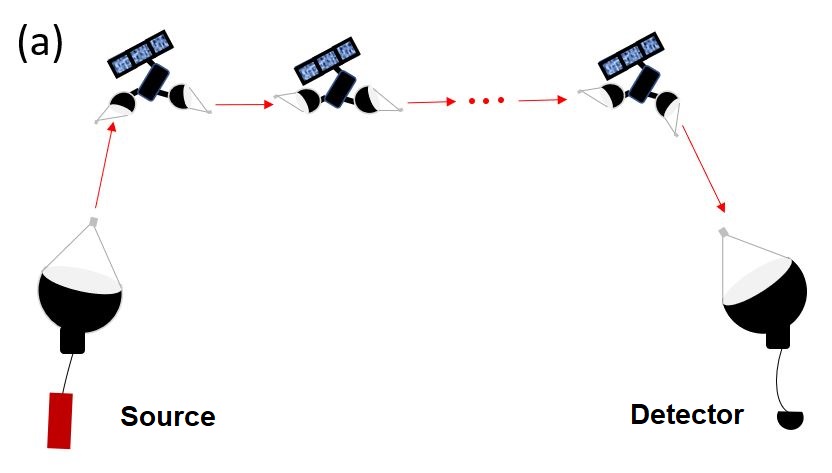}
    \end{subfigure}
    \begin{subfigure}{}
      \centering
      \includegraphics[width=.9\linewidth]{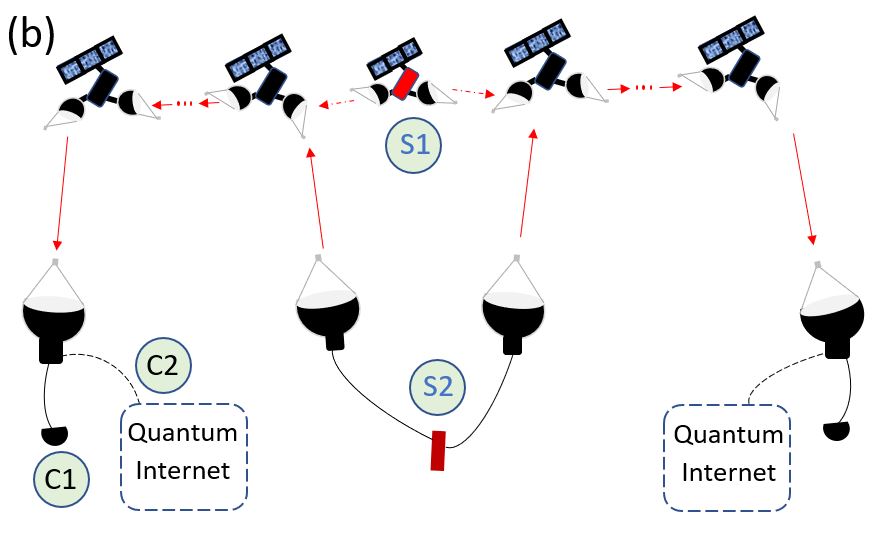}
    \end{subfigure}
    \caption{Different quantum communication schemes using a chain of reflector satellites, i.e. using All-Satellite Quantum Network (ASQN). (a) Qubit transmission – Photons transmitted from source in one ground station is reflected through space by a chain of reflector satellites to another ground station while beam diffraction is controlled by focusing. The three dots imply many satellites in between (not shown here). (b) Entanglement distribution -  Two options each for source and collection are shown. Two alternatives are shown for placing an entanglement source, one on satellite (S1) and one on ground (S2). Photon transmission and control of beam diffraction is similar as in (a). Transmitted entangled photons can be collected in ground stations in a detector (C1), e.g. to perform QKD or Bell inequality violation test, or they can be collected (C2) for further analysis  (e.g. building a quantum internet, as shown in Fig. \ref{QI_Appendix_Fig} in Section \ref{QI_appen})
    }
    \label{fig_scheme}
\end{figure}

Sending a photonic qubit from one place in earth to another through space has two issues and, in some sense, and they are intertwined. One is diffraction loss, and the other is the curvature of earth. Diffraction loss is itself not a limiting problem as the total diffraction loss between two 1 m diameter telescopes separated by 20,000 km in space is around 29 dB.  Although diffraction loss is not prohibitive when only two telescopes are used, the requirement of bending the light along the curvature of Earth quickly increases diffraction loss. To emit light from one point in earth, transmit through space and detect it at another point in earth we need to use multiple reflectors in space to guide the light along the curvature of earth. These reflector satellites would in turn cause more diffraction due to beam truncation at each satellite and subsequently even greater beam divergence. This would cause enormous diffraction loss which can only be stopped by decreasing beam truncation which would reduce both beam truncation loss itself and more importantly further beam divergence stemming from the truncated beam. There can be two ways of achieving this - either employing a small number of satellites with very large telescopes (diameters in several meters) separated by long distances or a large chain of satellites with smaller telescope (diameters in 40-60 cm) separated by a much smaller distance.

The latter path is chosen in this work as large telescopes get very heavy, difficult to manufacture and expensive. Hence, it would be rather difficult to build many large telescopes (needed to build a 2D network of satellites), fit them into spaceships and launch them to space. However, use of large telescopes may be still worth exploring more in future works to see if it has any unique benefits.

Each satellite behaves as a reflector, so that only mirror reflection contributes to photon absorption loss. However, the curvature in the telescopes is used to focus the light (as explained before) turning the satellites into effective ‘satellite lenses’. The chain of these ‘satellite lenses’ eliminates diffraction loss as shown by numerical modelling in Section \ref{result_simulation}. This forms the core of our protocols. The different protocols described are qubit transmission (introduced in this work), entanglement distribution and quantum internet (which requires quantum memories and is described in Section \ref{QI_appen}). Qubit transmission and entanglement distribution protocols are depicted in Fig. \ref{fig_scheme}. The qubit transmission scheme is introduced first, in Fig. \ref{fig_scheme} (a). As a protocol, it is conceptually the simplest. Photons are sent from the quantum source in the ground station, in the left, using a telescope towards the nearest satellite. The receiving telescope on the satellite collects the light and reflect it towards the other telescope, on the same satellite. The second telescope then transmit the light to the next satellite, towards the destination. The two telescopes in the satellite together give an appropriate amount of focusing to the light so that beam divergence due to diffraction is minimized. There can be several possible structures fro these two telescope system as described in Section \ref{telescope_setups} (shown in Fig. \ref{Diss_Fig1} ). 

The next satellite collects those photons and send them to the one after that. This process carries on until the destination point arrives when the satellite there sends the photons towards the destination ground station (right of the picture), aiming its telescope downwards. Telescope in the ground station collects these photons and either detects it (as shown in Fig. \ref{fig_scheme}), say for performing QKD, or uses it for some other purpose. 

Qubit transmission protocol has both source and detector on ground and consequently has multiple advantages associated with it. Possible advantages include ability to device and perform new experiments (e.g. sending or teleporting different quantum states like continuous variable states, squeezed states, multiphoton states etc.) by simply visiting the source and detector ground stations instead of doing new satellite launches, easy incorporation of future development and maintenance of source and detector, possibly larger frequency multiplexing capabilities due to abundance of physical space and electric power in ground station, ability to use large cryogenic coolers and other sophisticated techniques and other probable long term operational advantages. These may not be possible when the source or detector is on-board a satellite due to lack of access, lack of resources or simply lack of space. Even if possible on a satellite, many of these capabilities may be prohibitively expensive.

However, qubit transmission will have to face one uplink and one downlink transmission, instead of two downlink transmissions faced by entanglement distribution from a satellite based source. Uplink transmission has much larger atmospheric turbulence loss compared to downlink. Hence, qubit transmission faces a much larger effect of turbulence. Turbulence will increase the beam size many folds, resulting in greater total loss and diminished transmission rate. Turbulence loss depends critically on two factors. Increased propagation length after the atmosphere ends ($\sim$ 20 km from ground) and small receiving telescope size on-board satellite increases turbulence loss. However, in ASQN we inherently use low elevation ($\sim$ 200 km) satellites fitted with moderately big telescopes (40-60 cm), which will decrease turbulence loss. The effects of turbulence loss and the possible effects of turbulence in wavefront and hence in diffraction are calculated and discussed in Section \ref{result_turb}. 

In qubit transmission, one may use weak coherent pulse (WCP) as sources to perform decoy state QKD \cite{lo_decoy_2005}. WCP sources are easy to develop and have high rates. Another potential difference is frequency multiplexing. Although frequency multiplexing can be achieved for entanglement distribution too \cite{vergyris_fibre_2019, wengerowsky_entanglement-based_2018, pseiner_experimental_2021}, there can be limitations of doing so in ASQN - in a relatively small orbiting LEO satellite - in terms of physical space and possibly other resources (like electric power). In qubit transmission, the source stays in ground station and hence doesn't face these issues. This can probably pave the way for building larger frequency multiplexing capabilities in the free space based ASQN protocol. The advantages in using WCP sources and possible large multiplexing capabilities may compensate for the extra loss due to turbulence in qubit transmission. 

Other than qubit transmission, the protocol in Fig. \ref{fig_scheme}(a) can also perform entanglement distribution although with a quantum memory. One needs to simply store one photon from an entangled pair into a quantum memory and send the other one using the setup of Fig. \ref{fig_scheme}(a). However, this will need a quantum memory with long storage time and large multimode capacity. Although high memory efficiency is not needed as the entanglement distribution rate will drop simply by the factor of efficiency, developing such a quantum memory is still a difficult task. 

Entanglement distribution protocols that do not require a quantum memory are shown in Fig. \ref{fig_scheme}(b). An entangled pair of photons is sent along two directions from roughly around the mid-point between the two places where entanglement need to be transferred. The entangled photon source can be placed either in satellite or in the ground, S1 and S2 in Fig. \ref{fig_scheme}(b) respectively. The source satellite (S1, shown in red in Fig. \ref{fig_scheme}(b) ) is just a reflector satellite which carries a entangled source. The source can be removed from the beam path remotely, turning the source satellite into a reflector satellite. 

Diffraction loss in space (i.e., not considering ground links) is similar in the entanglement distribution protocol and in the qubit transmission protocol (Fig. \ref{fig_scheme}(a)) as diffraction is controlled by effective ‘satellite lenses’ in a similar way. The major difference in loss is in the ground links, especially due to uplink turbulence. Uplink turbulence is faced once with qubit transmission, none with entangled source in satellite (S1) and twice with entangled source in ground (S2). So, placing the entangled source in the satellite causes much smaller loss than with the source on the ground. However, even with larger loss a ground source can still be an interesting alternate option due to the ability to readily design new experiments involving different forms of entangled states and larger multiplexing capabilities.

So far, we have depicted proposals to accomplish qubit transmission and entangled distribution in the global scale without using quantum repeater protocol or quantum memories. Such a network can implement applications like fully secure global communication through QKD, where it is enough to simply detect the photons. However, to perform more sophisticated operation - like interfacing quantum computer for distributed computing - i.e., for full quantum internet capabilities quantum memories may be necessary and repeater protocols can be helpful. Different Quantum internet protocols built on the ASQN scheme and the requirements for different components are discussed in Section \ref{QI_appen}.

\section{Diffraction Analysis}\label{Results}

Quantum communication rate depends on the source rate and loss in the channel. Total loss in ASQN consists of beam divergence loss due to diffraction, loss through air transmission and individual satellite losses. Air transmission loss consists of loss due to atmospheric absorption and turbulence while individual satellite loss include reflection loss, beam pointing error, focal length error etc. In some of the cases diffraction loss and the satellite loss are interdependent, e.g. beam pointing error leads to diffraction loss. We first discuss beam propagation and diffraction loss which was the dominant loss in all the previously proposed long distances schemes through space and controlling which forms the core of ASQN. All through this work, we talk about photon propagation although photon and beam are used interchangeably at many places. Consequently for beam transmission, fraction of total intensity is used interchangeably to single photon transmission probability.

Each satellite has a system of telescopes to collect and transmit the photons towards the next satellite by reflection. A curved telescope mirror can be equivalently considered as a lens. Hence, the system of telescopes on a satellite together is modeled as a system of lenses which can further be represented by one effective lens, the 'satellite lens'. The focal length of that effective 'satellite lens' depends on the separation and the individual focal lengths of the telescope mirrors. As a special case the effective 'satellite lens' can have infinite focal length in which case the telescope setup in the satellite acts only as an aperture. Many different types of telescope systems can be used in our setups like on-axis and off-axis telescopes with different designs. Detailed analysis of these telescope systems is shown in Fig. \ref{Diss_Fig1} in Section \ref{telescope_setups}.

A chain of these 'satellite lenses' can efficiently control beam diffraction. Each 'satellite lens' has diameter $d$ and uniform separation between them $L_0$ as shown in Fig. \ref{theory_arg_fig} and Fig. \ref{Result_Fig1}(a). The lenses also have specific focal lengths. Both a larger separation L$_{satellite-ground}$ (or $L_{sg}$) and a larger telescope diameter d$_{ground}$ is considered at the end (in Fig. \ref{Result_Fig1}(a)) to account for the satellite ground link. Beam propagation and the consequent diffraction loss through this chain of satellites is investigated both numerically and theoretically. Before delving into the numerical calculations, a simple theoretical analysis is presented through Fig. \ref{theory_arg_fig}. For the theoretical analysis, beam truncation due to the finite lens size is ignored. We consider infinite sized lenses and simply investigate the change in beam size over distance. Although ignoring beam truncation sounds like an ideal case, truncation effects do become negligibly small around $ w_0 \sim d/4 $. Fraction of light intensity contained within a circle of radius $r$ of a Gaussian beam is given by $(1 – e^{-2r^2/w_0^2})$. Hence, almost all (99.96$\%$) the intensity is contained within $r = 2 w_0$. 

\subsection{Theoretical Arguments}\label{result_theory}

\begin{figure}
\includegraphics[width=\linewidth]{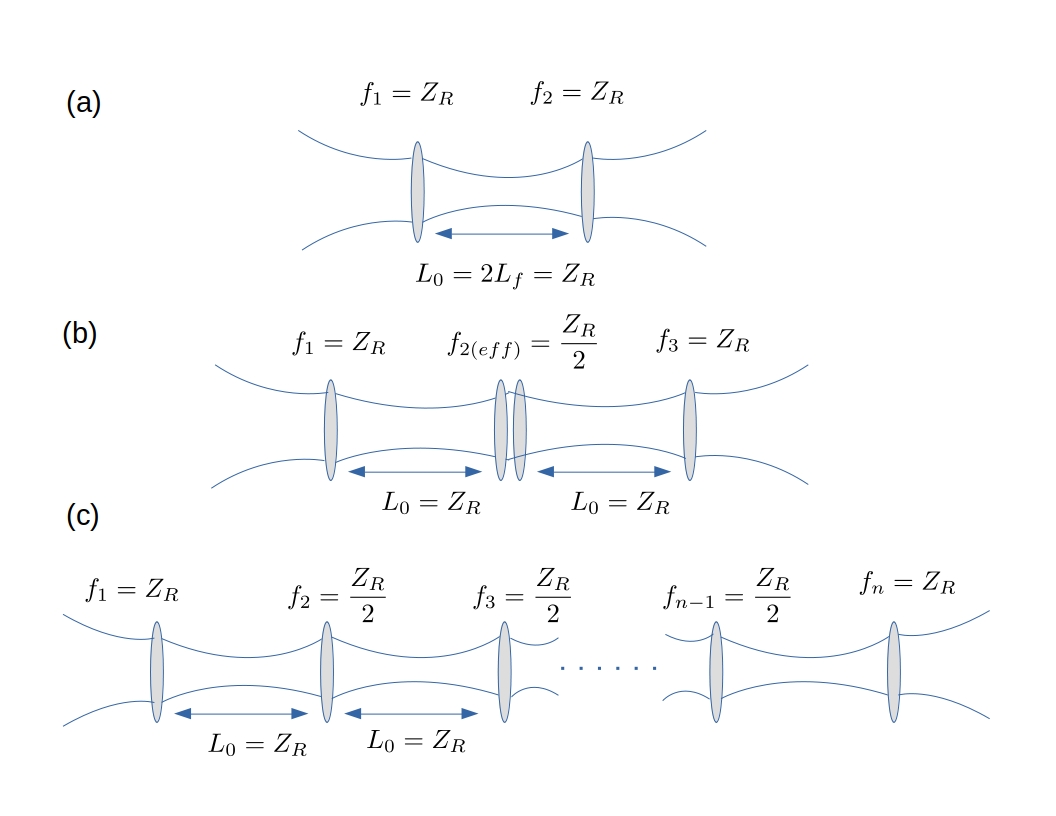}
\caption{Schematic diagrams explaining how a particular lens setup can completely contain beam divergence due to diffraction indefinitely. This is achieved with a uniform lens separation as large as the Rayleigh range ($z_R = \pi w_0^2/\lambda$) }
\label{theory_arg_fig}
\end{figure}

For non-truncating lenses, we would show that for small enough lens separation ($L_0$), beam diffraction can be completely controlled by choosing a proper set of focal lengths. For simplicity, we consider the case of an incident Gaussian laser beam with beam waist ($w_0$) at the position of the first lens. It is not essential for the beam to be incident at the beam waist, and it is only assumed for convenience in the calculations, which is explained in detail later. For a Gaussian beam with wavelength $\lambda$ and beam waist $w_0$, Rayleigh range $ z_R$ is defined as $ z_R = \pi w_0^2/\lambda$. We would show that diffraction is completely controlled for a lens system with lens separation $L_0 = z_R $ and focal lengths ($f$) of successive lenses as $z_R, z_R/2, z_R/2, \cdots $. Such a system would completely preserve the beam size and wouldn't cause any beam divergence indefinitely (i.e even at infinite distance). This can be directly verified using the ABCD matrix formulation for Gaussian laser beams \cite{renk_basics_2012}. We explain it more intuitively below, although along the same lines.

This is explained in three stages in Fig. \ref{theory_arg_fig}. If a Gaussian laser beam is focused by placing a lens (with focal length $f_1$) at the beam waist and another lens, with focal length $f_2 = f_1$, is placed at double the focusing distance ($L_f$ with $L_f$ = $L_0/2$) then the beam will return to its beam waist after the second lens which is evident from symmetry of the setup. Note that the focusing distance $L_f$ is not the same as the lens focal length (i.e. $f_1$) because the input beam is a Gaussian laser beam and not a constant wave-front. As the beam is at beam waist, its original input state for Fig. \ref{theory_arg_fig}(a), placing another set of lenses with focal length $f_1$ and double focusing distance (i.e. 2$L_f$) separation (i.e. exactly as in Fig. \ref{theory_arg_fig}(a)) would result in the beam returning to again to its beam waist another 2$L_f$ distance apart (a shown in Fig. \ref{theory_arg_fig}(b)). This process can be repeated indefinitely to make the beam return to its beam waist at indefinitely far away distances. The effective focal length of the middle lenses would be $f_1/2$ - effective focal length when combining two lenses of focal length $f_1$. Hence, a lens configuration of $f_1 = 2 f_2 = 2 f_3 = …= 2 f_{n-1} = f_n $ would indefinitely contain beam divergence when placed at uniform separations of $L_0$ = 2$L_f$, as shown in Fig. \ref{theory_arg_fig}(c).

The maximum possible lens separation for our setup - $L_0$ = $z_R$ - can be found as following. When a Gaussian beam with Rayleigh range $z_R$ is focused by a lens of focal length $f$ placed at beam waist, the beam is focused at a focusing distance $ L_f = f/(1+f^2/z_R^2)$, which is different from the lens focal length $f$ \cite{renk_basics_2012}. The focusing distance $L_f $ is  maximum at $f = z_R $ with $L_{f(max)} = z_R/2$. If $ L_0 = z_R $, the beam is focused at the midpoint between the first and second lens ($L_{f(max)} = z_R/2$) and hence the beam profile is symmetric around the midpoint.

As stated earlier, It is not essential for the light to be incident at the beam waist on the first lens with focal length $ f = z_R$. For example, we may simply start from the second lens (i.e. consider the second lens as the first lens) with a divergent input beam. More generally, for the scheme to work the same light can be incident with any curvature with a beam spot same as the beam waist and focal length of the lens can be adjusted accordingly. This means the first lens can have a small focal length and it can capture an extremely divergent beam coming from an optical source, situated in the same satellite, as would be the case for entanglement distribution from satellite source (as shown in Fig. \ref{Result_Fig1}(a)).

In ASQN the effect of Earth curvature is not taken into account directly as we are considering the lenses in a straight line here, both for theory and simulation. However, we would argue that it does not affect the overall calculation significantly. Firstly, the angle of the light will be bent at each satellite (and hence at each ‘satellite lens’) is very small, less than half a degree for satellite separation of around 100 km. In Fig. \ref{Diss_Fig1} we showed how this angle can be given to the light beam using telescope mirrors. This is simply changing the angle of the beam without providing any extra focusing similar to how plane mirrors are used in optics experiments to change the angle of a laser beam. Hence, this extra angle would not have any effect on diffraction, i.e. it wouldn't change the beam waist.

\subsection{Simulation}\label{result_simulation}

\begin{figure}

\begin{subfigure}{}
  \centering
  \includegraphics[width=.9\linewidth]{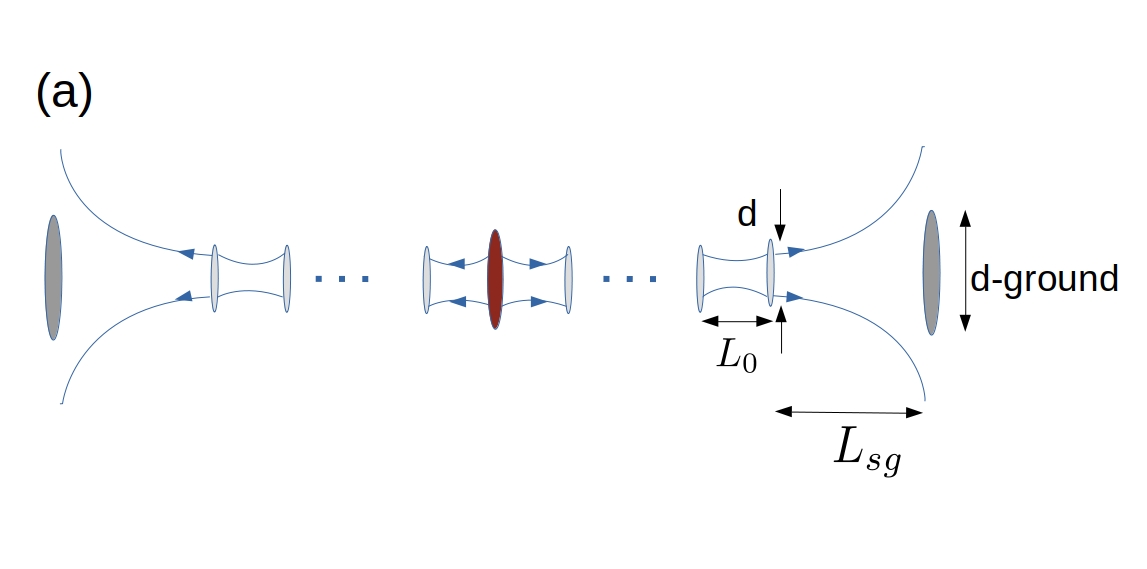}
  \label{fig:sub-first}
\end{subfigure}
\begin{subfigure}{}
  \centering
  \includegraphics[width=.9\linewidth]{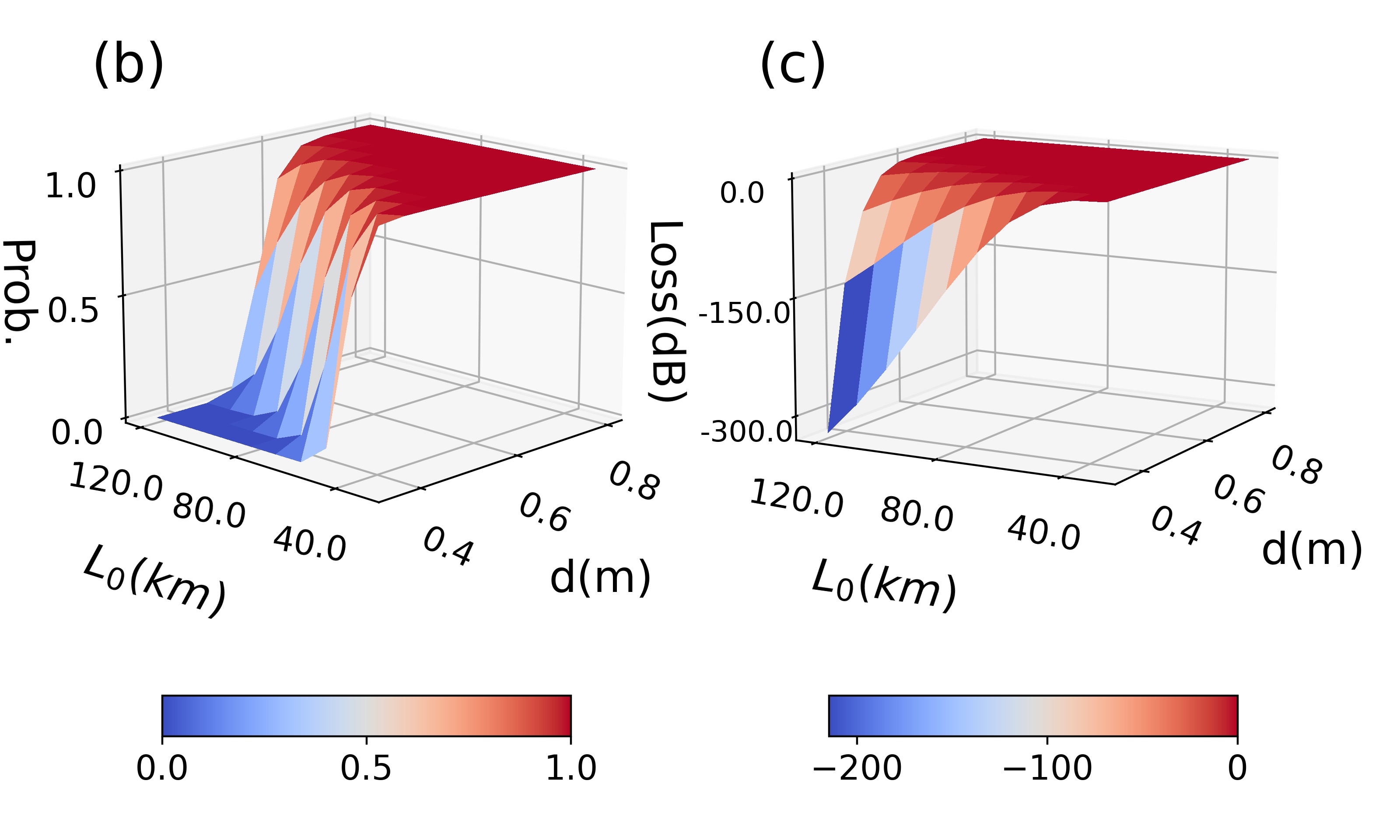}  
  \label{fig:sub-first}
\end{subfigure}
\begin{subfigure}{}
  \centering
  \includegraphics[width=.9\linewidth]{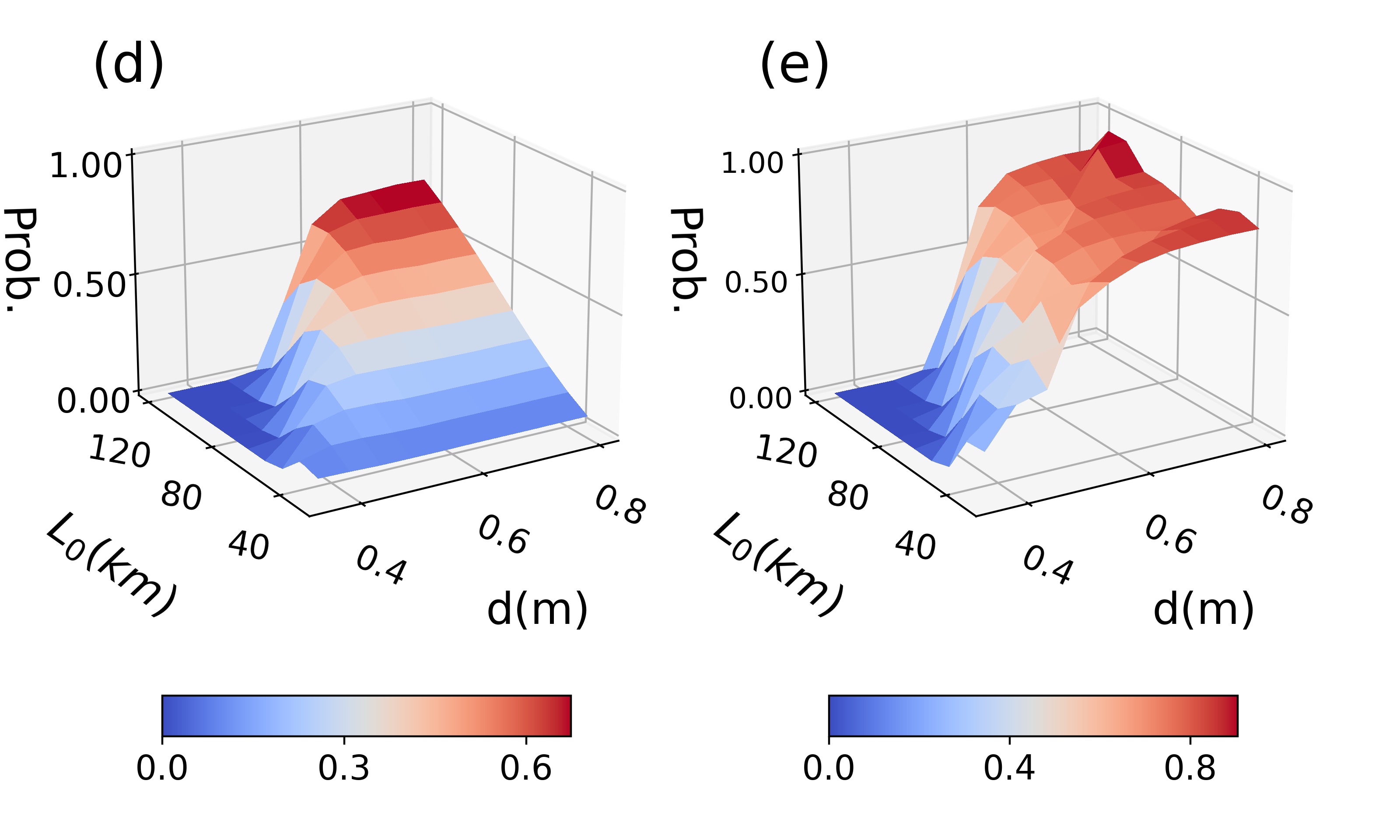}  
  \label{fig:sub-second}
\end{subfigure}
\caption{Diffraction loss in entanglement distribution. - (a) Schematic diagram showing the entangled pair source containing 'satellite lens' in the middle and the ground links at both the ends. The effective 'satellite lenses' contain diffraction by controlling beam divergence. (b) Numerical simulation of diffraction loss showing entanglement distribution probability at 20,000 km for different telescope diameters ($d$) and satellite separations ($L_0$), without considering ground link. (b) and (c) are essentially same graphs with different scales (regular and logarithmic) showing respectively low (b) and high (c) loss regions clearly. (d-e) Diffraction is simulated with ground links included. In (d) using ($f_{N-1}, f_N$) = ($L_0/2, L_0/2$) results in poor performance for small $L_0$ values. However, on optimising ($f_{N-1}, f_N$) a plot closer to the original one (b) is found in (e).}
\label{Result_Fig1}
\end{figure}

\begin{figure}
\begin{subfigure}{}
  \centering
  \includegraphics[width=.9\linewidth]{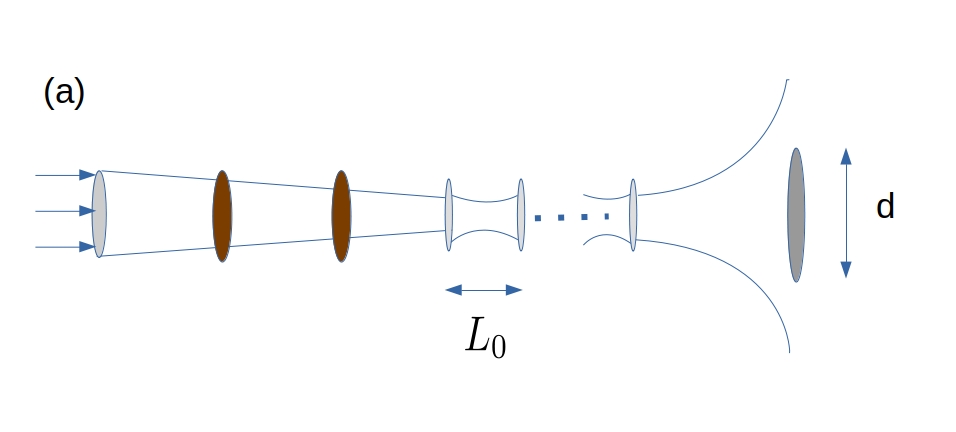}
  \label{fig:sub-first}
\end{subfigure}

\begin{subfigure}{}
  \centering
  \includegraphics[width=.9\linewidth]{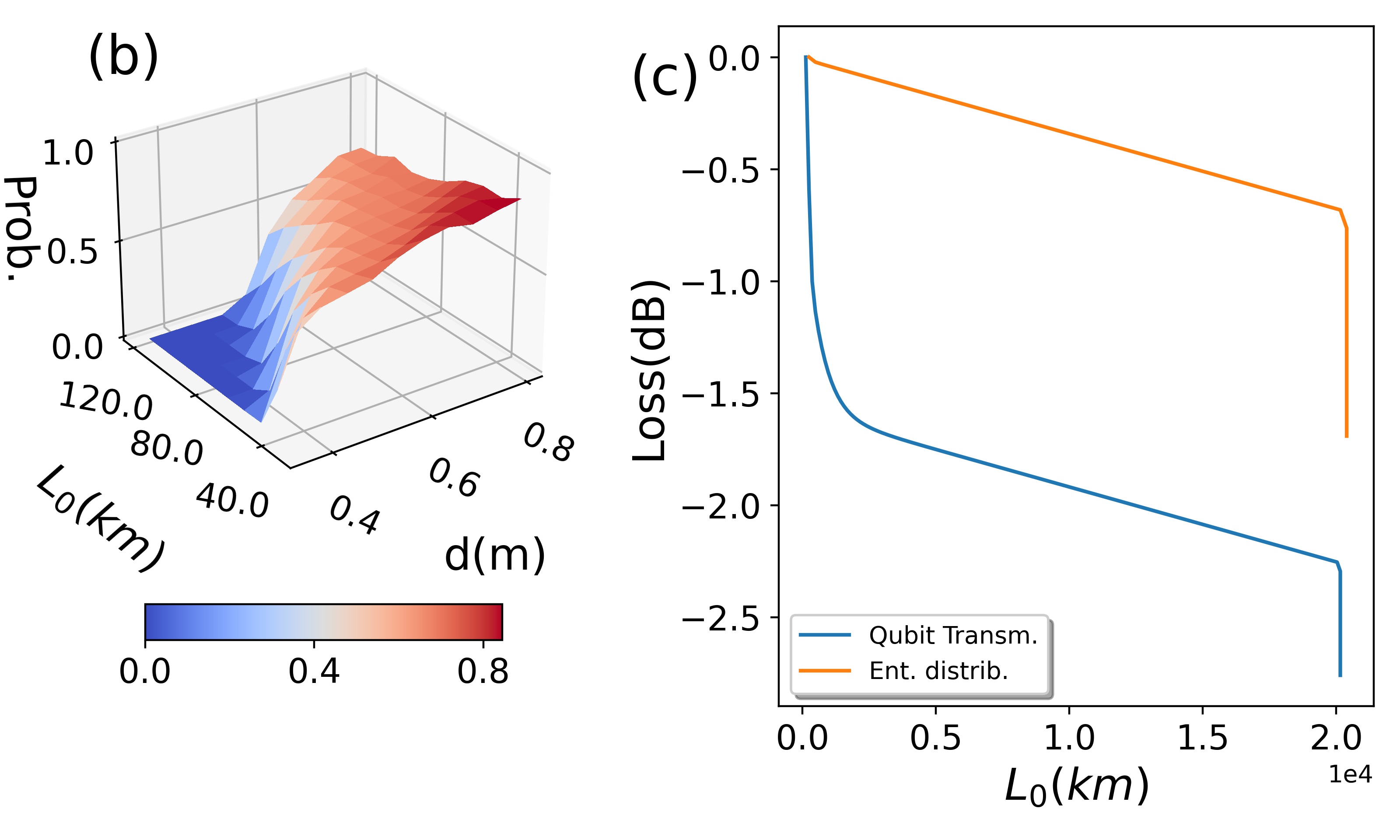}  
%   \caption{Put your sub-caption here}
  \label{fig:sub-second}
\end{subfigure}
    \caption{ Diffraction loss in qubit transmission - (a) Schematics of Qubit transmission protocol has been shown with lenses (grey colored) and apertures (brown colored) which are basically lenses with infinite focal length. The protocol here starts from the uplink receiving telescope and the ground link in the downlink is shown in the right. Please see the text for details. (b) Diffraction loss for qubit transmission at 20,000 km for different telescope radius ($d$) and satellite separation values ($L_0$) are shown. Ground links are not considered (neither the initial uplink nor the eventual downlink) while 800 nm light is used. (c) Photon transmission probabilities are shown with distance with $d$ = 60 cm and $L_0$ = 120 km for both qubit transmission and entanglement distribution. Both down links are considered for entanglement distribution while only the eventual downlink is considered for Qubit transmission.  Uplink loss in tens of dB is too large to see any comparison. Hence, uplink loss is included only in Fig. \ref{Result_Fig3}(e)-(f). Initially probability falls quickly for qubit transmission due to some beam truncation and diffraction at the apertures, although this produces less than 2 dB  additional loss.}
    \label{Result_Fig2}
\end{figure}

We simulate diffraction loss numerically next, including beam truncation. We first discuss the case for entanglement distribution with the source on satellite (S1 in Fig. \ref{fig_scheme}(b)). The entanglement distribution setup considered for the simulation is shown in Fig. \ref{Result_Fig1}(a). The source satellite is shown in the middle, while the ground links are at both ends.

Diffraction loss at the global distance of 20,000 km for different values of telescope (lens) diameter $d$ and satellite separation distance $L_0$ are plotted in Fig. \ref{Result_Fig1}(b) for photons of wavelength 800 $nm$. To perform the simulation, for each different $L_0$ We considered a beam waist of $w_0 = \sqrt{L_0\lambda/\pi}$. Then a lens system $L_0$, $L_0/2$, $L_0/2,\cdots $ is used to contain the beam diffraction, as argued in Section \ref{result_theory}. If the lens diameter $d$ is big enough to contain the beam, there would be no loss at all due to diffraction as understood theoretically. For smaller lenses (i.e. telescopes), the beam will get truncated which would lead to loss and for significant truncation this would result in beam divergence due to diffraction which would further increase the loss very rapidly. As an example, for $L_0$ = 120 km we have only 0.67 dB loss for $d$ = 60 cm when only $0.28\%$ of the beam is truncated by the telescope while for $d$ = 35 cm loss becomes an enormous 324 dB as $13.47\%$ beam truncation occurs. Any loss other than diffraction (e.g. turbulence loss) is not included here.

Fig. \ref{Result_Fig1}(b) shows transmission probability of an entangled pair of photons at a distance of 20,000 km. To produce this graph we simulated photon transmission to a distance of 10,000 km, without considering the final ground link, and then squared it to account for the pair. Fig. \ref{Result_Fig1}(b) clearly establishes that in our protocol for a certain combination of 'satellite lens' diameter and the distance between two satellites we can achieve transmission with very minimal loss. We see the expected result that diffraction loss is highest for small $d$, large $L_0$ values while it is least for large $d$, small $L_0$.  Diffraction can be seen to be constant in places where $L_0$ is proportional to $d^2$. This is also expected as $ L_0 = z_R = \pi w_{0}^2/\lambda$ and diffraction loss in ASQN depends on the ratio of $w_0$ and $d$. In Fig. \ref{Result_Fig1}(b)  the region of low loss is clearly seen. The exact same plot for diffraction loss  is expressed in units of decibel in Fig. \ref{Result_Fig1}(c), which clearly shows the region with high loss. At this distance of 20,000 km, loss is seen to be as high as 324 db for the smallest lens diameter chosen as 35 cm and largest satellite separation $L_0$ of 120 km.

\begin{figure}
\begin{subfigure}{}
  \centering
  % include first image
  \includegraphics[width=.9\linewidth]{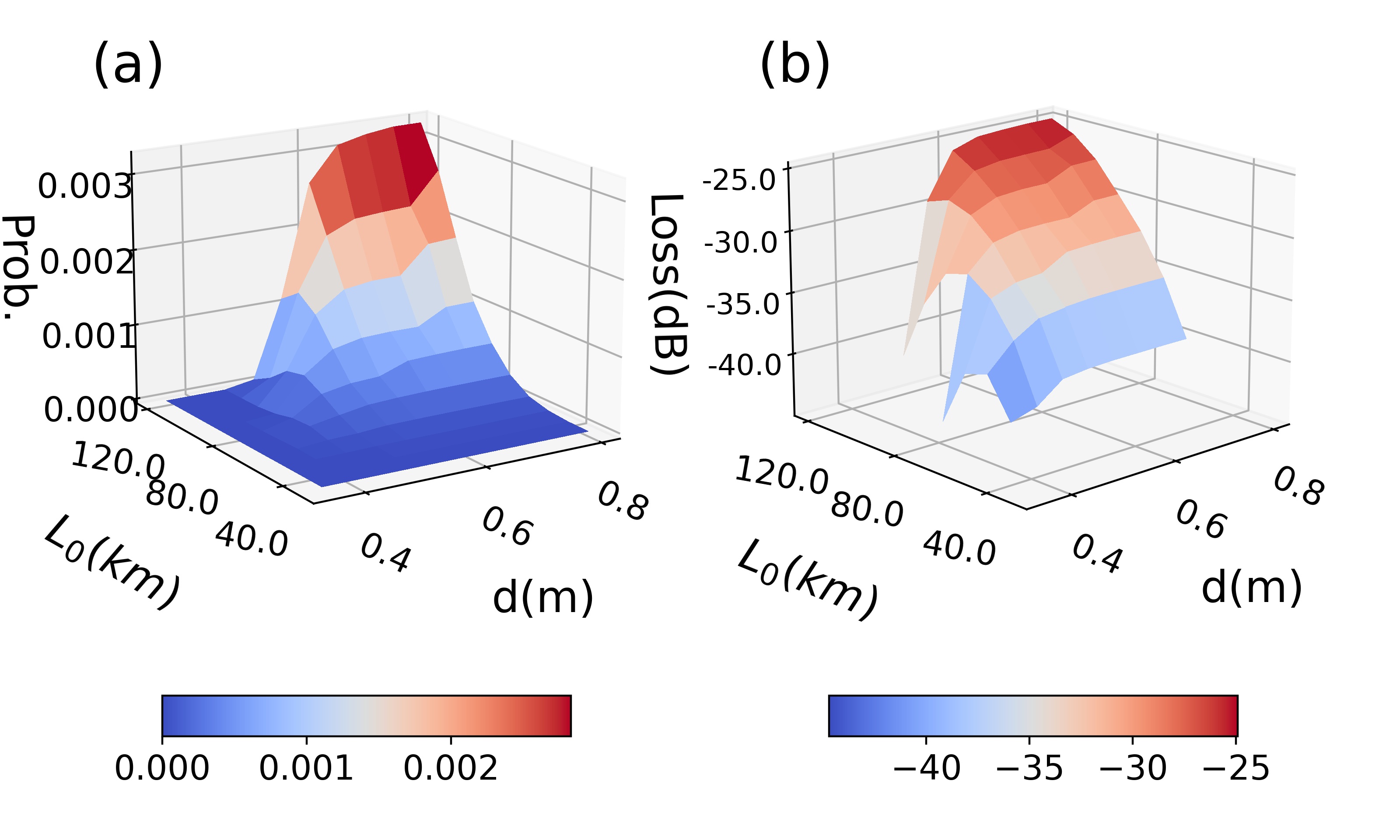}  \label{fig:sub-first}
\end{subfigure}

\begin{subfigure}{}
  \centering
  % include second image
  \includegraphics[width=.9\linewidth]{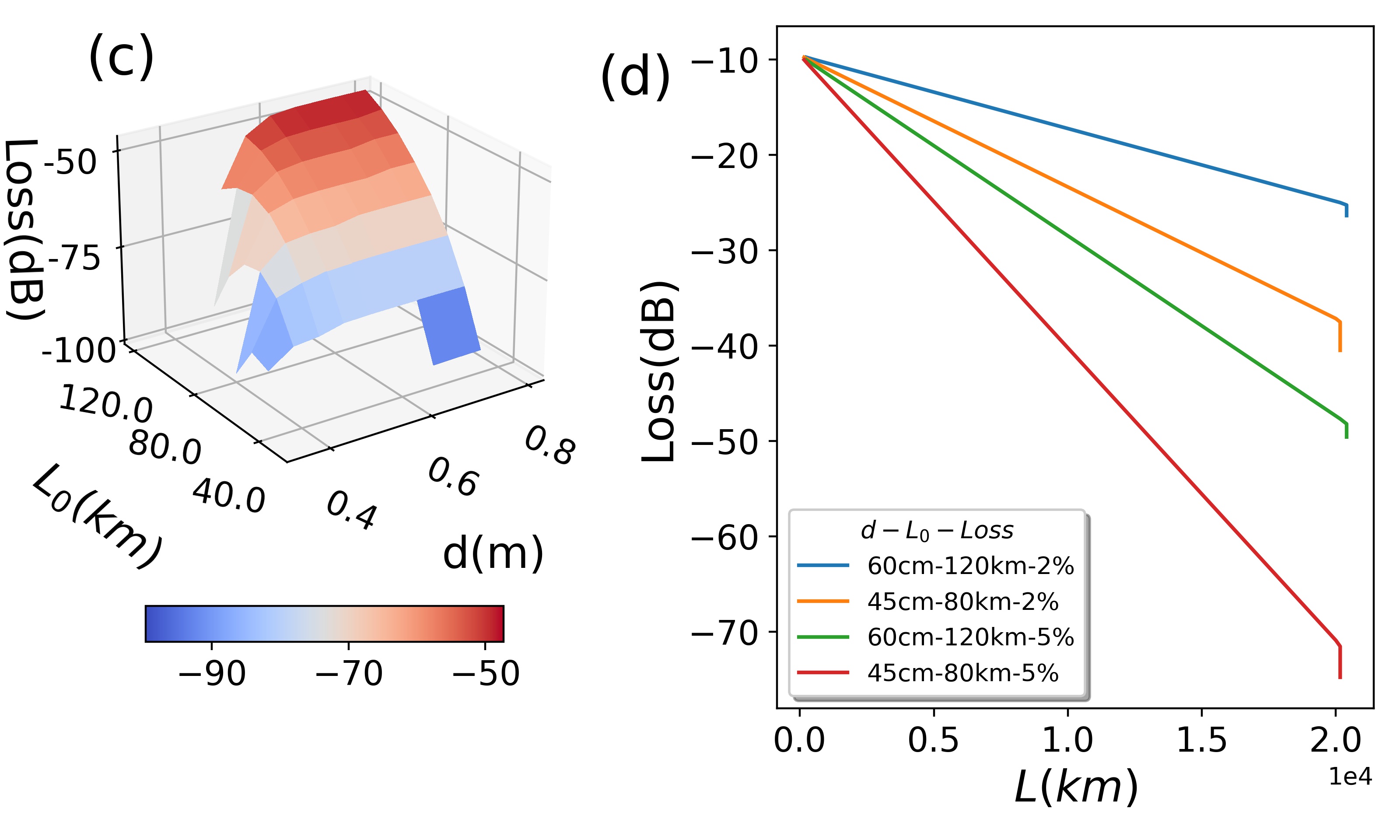}  
%   \caption{Put your sub-caption here}
  \label{fig:sub-second}
\end{subfigure}
\begin{subfigure}{}
  \centering
  % include second image
  \includegraphics[width=.9\linewidth]{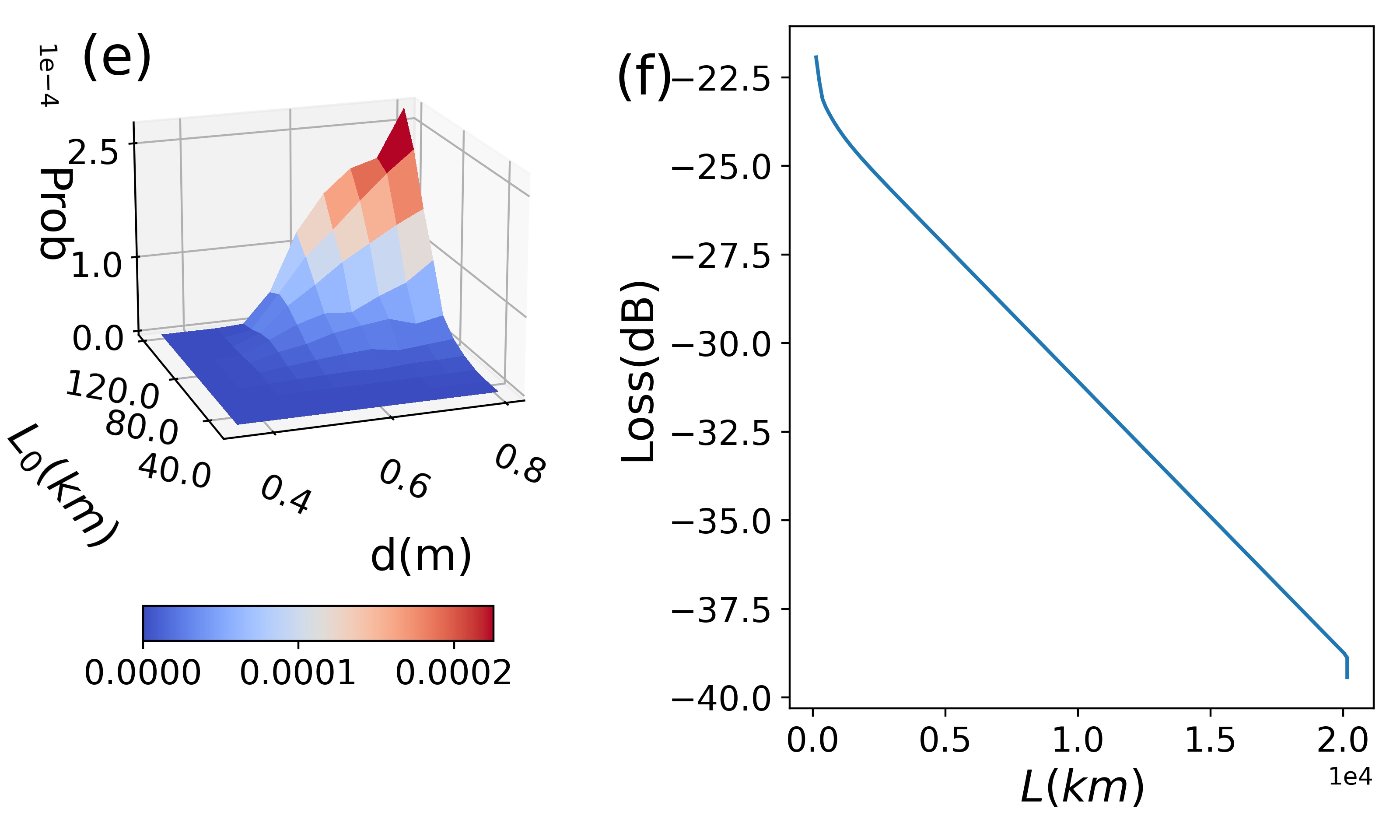}  
%   \caption{Put your sub-caption here}
  \label{fig:sub-second}
\end{subfigure}
% \begin{figure}[htbp]
%     \centering
%     \includegraphics[width=0.95\linewidth]{Result_Fig3.jpg}
    \caption{Total loss in photon propagation – (a-b) Total loss for entanglement distribution with 2$\%$ absorption loss for each satellite. Same graph is shown in (a) and (b) with (b) showing loss in decibel scale. Only points with loss up to 45 dB is shown in (b) to clearly show the loss values for smaller $L_0$ values.  In both graphs there exist a minimum loss (at an optimum $L_0$) for each $d$ value, either occurring inside the plotted region or outside. (c) Same plot as (b) with 5$\%$ satellite loss and showing points up to 100 dB loss. (d) Entangled distribution loss is shown with distance for different ($d$, $L_0$) values and diffraction loss values (satellite loss 2$\%$ of 5$\%$) for a total propagation of 20,000 km. (e) Qubit transmission protocol total loss values (including both uplink and downlink loss, satellite loss of 2$\%$ and other losses) are shown for different ($d$, $L_0$) values. (f) Qubit transmission loss is shown with distance for $d$ = 60 cm, $L_0$ = 120 km and 2$\%$ satellite loss.}
    \label{Result_Fig3}
\end{figure}

In Fig. \ref{Result_Fig1}(d) we add the ground link to estimate the total diffraction loss. For the simulation, the diameter of the ground telescope used is taken as 60 cm whereas the satellite elevation is taken as 200 km.  Here we have used the same lens configuration as before( $L_0, L_0/2, L_0/2,\cdots $), i.e. all lenses after the first lens, including the last two lenses before ground link  $f_{N-1}$ and $f_N$, have focal length $L_0/2$. This resulted in the region of maximum intensity to shrink. The region with small $L_0$ values has their probability diminished as the above lens configuration diverges beams more at small $L_0$ and then they have to travel 200km distance in the ground link. However, this issue is not insurmountable. In Fig. \ref{Result_Fig1}(e)  we tried this by optimizing the lens configuration (by changing $f_{N-1}$, $f_N$ values) to increase the area of maximum intensity in the plot. Here we have adjusted the focal length of the last two lenses before ground link, so that the intensity at the ground telescope is maximum. It should be noted that even without optimization, in Fig. \ref{Result_Fig1}(d) , intensity did not decrease a lot in case of large $L_0$ values which is really our desired regime, as we would see later in Fig. \ref{Result_Fig3} while considering total loss. Hence, this regime can function even without adjustable focal lengths.

For qubit transmission with both source and detector on ground (Fig. \ref{fig_scheme}(a)), diffraction loss would be similar to entanglement distribution if effects of atmospheric turbulence can be completely neglected. However, as discussed before the source on ground means photons would have to face atmospheric turbulence at first (or uplink turbulence) which results in huge beam divergence when photons reach the satellite \cite{bonato_feasibility_2009}. Considering the large beam spot at satellite, the portion of the beam captured by the satellite telescope can be considered as a constant wavefront. Although this is not completely true as the beam gets fragmented due to turbulence \cite{villasenor_enhanced_2021}, such an analysis would still produce quite an accurate result as shown in Section \ref{result_turb}.

Diffraction due to such a constant wavefront can be controlled using a two-step strategy. Firstly, the constant wavefront can be focused using the first lens on a later lens (not necessarily the very next one) as shown in Fig. \ref{Result_Fig2}(a). As a constant wavefront is focused by a lens, due to Fraunhofer diffraction an Airy disk pattern would be formed at the focus \cite{born_principles_1999}. The Airy disk intensity profile is a complicated expression depending on Bessel function. However, most of the light energy is contained within the central disk ( $\sim$ 84$\%$ ), i.e. inside the first minima of diffraction \cite{born_principles_1999}. This central peak in itself looks similar to a Gaussian and hence it can be reasonably expected the Airy disk pattern would propagate somewhat like a Gaussian beam with similar sized beam waist. The scheme is designed based on this assumption and the simulation results justify it. The size of the central Airy disk at the focus is given by $w_{Airy} = (n L_0 1.22 \lambda)/d$, where $n$ is the number of lenses after which the beam is focused ($n$ = 3 in Fig. \ref{Result_Fig2}(a)). The number $n$ is chosen such that the size of Airy disk is larger than a corresponding Gaussian beam waist with Rayleigh length $L_0$, i.e. $ w_{Airy} > w_0 = \sqrt{L_0\lambda/\pi}$. This would ensure that beam diffraction can be controlled using 'satellite lenses'. The lenses between the first lens and the target lens (i.e $n^{th}$ lens) would be removed, i.e. they would simply act as apertures of diameter $d$. As we consider the Airy disk pattern at the focus similar to a Gaussian at beam waist, our original lens configuration of $L_0$, $L_0/2$, $L_0/2, \cdots$ can be employed for the following lenses, starting at the $n$th lens.

Fig. \ref{Result_Fig2}(b) shows numerical simulation of diffraction loss for qubit transmission at global distances of 20,000 kilometers at different d, $L_0$ values. The simulation follows the setup of Fig. \ref{Result_Fig2}(a)  without considering the ground link. In contrast to the entanglement distribution case, in qubit transmission light transmission probability does not reach unity asymptotically even while not considering ground link. There can be several reasons for this. One of them is the Airy disk beam profile which is not actually a Gaussian and hence would cause some errors on propagation. Another possible reason would be the diffraction effects of the initial apertures (where the lenses are removed) on the light beam. Due to the combined effect of these factors, light intensity drops in the initial propagation of the beam before stabilizing. This is seen in the Fig. \ref{Result_Fig2}(c) where light intensity is plotted at different lengths for the case $d$ = 60 cm, $L_0$ = 120 km, $\lambda$ = 800 nm for both entanglement distribution and qubit transmission. The total intensity captured at each lens is calculated to plot this graph. Any loss other than diffraction loss (e.g. turbulence loss) is not included here. In both cases, ground link is included after the last two lenses and focal lengths $f_{N-1}$ and $f_N$ were optimized for the ground link transmission. Other than the initial fall discussed above, intensity decreased in similar fashion for qubit transmission as it is for entanglement distribution.

The lens arrangements (without considering ground links) for both entanglement distribution ($L_0, L_0/2, L_0/2,\cdots $) and qubit transmission ($n L_0,\infty ,\infty , \cdots ,L_0, L_0/2, L_0/2,\cdots $) to contain diffraction are just special lens configurations. These lens configurations work ideally only for certain cases, e.g. the case we described in the theory for the entanglement distribution where $d$ is so large that there is no beam truncation. However, this lens arrangement is not the best for a general $d, L_0$. An optimization over the whole lens configuration or even a part of it (i.e. considering only a few lenses) can probably improve the diffraction loss, at least in the cases where loss is starting to increase. We may have a much smaller maximum loss than the $\sim$ 300 dB loss seen in Fig. \ref{Result_Fig1}(c), using an optimized lens configuration. One may optimize over just a few lenses (say ten lenses) such that the light is confined within them and at the end the initial beam profile is returned. For subsequent lenses this configuration is repeated to confine the light. In general, such configuration of lenses to confine light is called lens waveguide system in the literature \cite{siegman_modes_1967}. For example, a lens waveguide system to confine light along a curved path is shown in \cite{marcuse_propagation_1964}.

We now discuss losses other than the diffraction loss. One of these is air transmission loss which consists of two parts - absorption in air and atmospheric turbulence. These losses are only present in the satellite-ground links. They are not factors in satellite-to-satellite transmission as air density and hence absorption loss becomes precipitously low in high elevations. Atmospheric absorption loss in the ground link depends heavily on the optical frequency and the angle of transmission. Absorption losses increase exponentially with distance and hence grazing incidence is lethal for quantum information transfer. In our scheme, due to the closely spaced satellites we don't reach a high angle of incidence.

% For example, in 200 km elevation absorption loss through 20,000 km of air transmission is only …… loss.

Air turbulence contributes to loss in satellite-ground link in both uplink and downlink along with diffraction. Turbulent eddies in the atmosphere cause beam wander and beam spreading \cite{fante_electromagnetic_1975}. Turbulence is much more in the uplink than downlink transmission, as in uplink the dephased beam emanating from the turbulent atmosphere (which ends at around 20 km) has to travel a long way to reach the satellite where it spreads a lot more due to diffraction. In the downlink however there is no propagation after the atmosphere and hence turbulence loss is much lower. Because the loss is due to the large propagation distance in the uplink it can be reduced simply by allowing less propagation distance, i.e.  using low elevation satellites. This is anyway native to ASQN because of the existence of the satellite chain. Due to the small separation ($\sim$ 100 km) between satellites in the chain, a large field-of-view is not required for each satellite and hence satellites don't need to be at high elevation. Low elevation satellites also decrease diffraction loss in ground link. Turbulence loss can be decreased further using larger diameter receiving telescopes and shrinking the initial beam waist to limit spreading. The effect of turbulence will fragment the beam. However, the beam can still be focused. Further modelling of the turbulence related loss and beam propagation effects are done in  Section \ref{result_turb}.

The other major contributor to photon loss is termed as satellite loss which encompasses all losses caused by one satellite while reflecting the photon towards the next one. This loss has two parts essentially – the loss intrinsic to the satellite (e.g., the mirror reflection loss) and satellite errors which essentially worsens the diffraction loss. The intrinsic loss grows exponentially with the satellite number and hence must be controlled at a very small value (or the number of satellites needs to be reduced dramatically). Reflection loss at each satellite mirror is the most significant of the exponential scaling losses, especially more so as there are multiple mirrors needed on one satellite itself. Two kinds of reflectors can be used - standard metal reflectors and Bragg reflectors. Standard metal reflectors - gold or silver coated mirrors - have reflectivities of at most 99.5 $\%$, available only in wavelengths above 1 $\mu m$. More sophisticated dielectric mirrors or distributed Bragg reflectors or simply Bragg mirrors can have very high reflectivity by using multiple thin layers of different refractive index glasses. They can have refractive index as high as 99.9999 $\%$ and can be manufactured for almost any frequency. These mirror systems are discussed in detail later.

There can be several other factors contributing to the satellite loss. This includes mirror positioning error which results in focal length error, mirror angular position error which may cause beam deviation etc. The positioning error of the satellite itself would also contribute to loss. Moreover, till now we assumed monochromatic light for our calculation. So, the finite frequency width of a photon may also influence diffraction. The individual and the combined effects of all the above-mentioned issues are discussed in Section \ref{error_appen}. There are other losses in source or ground stations, like detection loss which has been considered too, details of which is given in Section \ref{simulation_method}.

We showed the effect of diffraction loss alone in Fig. \ref{Result_Fig1} and Fig. \ref{Result_Fig2}. Now, along with the diffraction loss (with ground link loss optimized) we also consider the loss due to every satellite, which is mentioned as exponential satellite loss above. Moreover, other losses - atmospheric absorption, turbulence, satellite chain setup error, detector loss etc. - were also considered that produced an constant overhead loss. With 2$\%$ exponential loss for each satellite in Fig. \ref{Result_Fig3}(a), for each satellite the total loss scales exponentially with the number of satellites. This is because diffraction loss due to truncation also scales almost exponentially when multiple satellites are used (as shown in Fig. \ref{Result_Fig1}(b)). That is why truncation loss can not be tolerated at all and must be kept very low. In Fig. \ref{Result_Fig3}(a) for a certain lens diameter ($d$) when the satellite separation ($L_0$) decreases, the number of satellites used for the whole propagation distance increases, resulting in exponential loss. Next in Fig. \ref{Result_Fig3}(b), we plot the same loss in decibel units up to a certain loss (45 dB) value. We see that the loss increases quickly with decreasing $L_0$ due to the increasing number of satellites required which contributes to satellite loss. This is already seen in Fig. \ref{Result_Fig3}(a) where probability is plotted directly. However, satellite loss doesn’t reach anywhere as high as the maximum loss due to diffraction (in Low $d$, high $L_0$ values, Fig. \ref{Result_Fig1}(c) ). Hence, we plot data points only upto a certain loss value (45 dB) to be able to show the effect of satellite loss distinctly. Later on in Fig. \ref{Result_Fig3}(c), we also show the effect of a much higher loss at 5$\%$ exponential satellite loss. Here as expected, much more loss occured for the same ($d$, $L_0$) values and hence loss values upto 100 dB is shown. In all three figures showing total loss (Fig. \ref{Result_Fig3}(a)-(c)) for a certain lens (i.e. telescope) diameter value $d$ we can find an optimum loss value at a particular $L_0$. The existence of optimum can be seen in Fig. \ref{Result_Fig3} (a)-(c) by looking at 2D sections of the 3D plot, where graphs peak in $L_0$ for fixed $d$ values - at different L$_0$ values for different $d$ values. The optimum occurs due to a trade-off between high diffraction loss at large $L_0$ and high satellite loss at small $L_0$. We have shown total intensity with propagation distance ($L$) at these optimum $d, L_0$ values in Fig. \ref{Result_Fig3}(d) for both 2$\%$ and 5$\%$ satellite loss. The $d$ values for which optimum isn't visible in the graphs (e.g. $d  >$ 55 cm for 2$\%$ loss) has optimum at even higher $L_0$. We have taken the lowest values of loss (at highest $L_0$ values) for these $d$ values in Fig. \ref{Result_Fig3}(d). In Fig. \ref{Result_Fig3}(d), best intensity scaling with distance is seen for ($d, L_0$ = 60cm, 120km) values for 2$\%$ satellite loss. However, even for much smaller $d$ values, reasonable intensity scaling is seen for (45cm, 80 km) for 2$\%$ satellite loss and (60cm, 120 km) for 5$\%$ satellite loss. Although there is more than 40 dB loss at full global distance of 20,000 km at the more intermediate distances ($\sim$ 10,000 km) there is much less loss. Even for the drastic case of ($d$ = 45 cm, $L_0$ = 80 km) for the 5$\%$ satellite loss the loss at $\sim$ 5,000 km is only 25 db. With a telescope size of 45 cm this would be still quite interesting as the best entanglement distribution record till date is only upto 1200 km \cite{yin_satellite-based_2017}. Although, there are several other loss mechanisms (like aberration, polarization aberration etc.) that are not accounted for here. We argue in Section \ref{Discussion} that these  either don't cause a significantly large amount of loss in our case or certain system designs can be incorporated to mitigate the effects of the loss. However, to confirm these effects a more detailed theoretical analysis of these loss mechanisms should be done along with experimental verification by doing tabletop experiments. Some of these experiments are suggested later in Section \ref{tabletop}.

Similar to entanglement distribution, the effect of total loss was shown for qubit transmission in Fig. \ref{Result_Fig3}(e).  The effect of 2$\%$ satellite loss was added to the diffraction loss with optimized ground link loss. The effect of satellite loss on qubit transmission is quite similar to that of entanglement distribution, seen in Fig. \ref{Result_Fig3}(a). The plot is slightly uneven which originates from the uneven diffraction loss (seen in Fig. \ref{Result_Fig2}(b)) and ground link optimization for qubit transmission. In Fig. \ref{Result_Fig3}(f), similar to Fig. \ref{Result_Fig3}(d) intensity is plotted with distance for (60 cm, 120km) values of ($d$, $L_0$) for $2\%$ satellite loss. This graph shows the same exponential trend of Fig. \ref{Result_Fig3}(d) along with the initial diffraction loss seen at the first few apertures for qubit transmission (Fig. \ref{Result_Fig2}(c)). The Initial overhead loss (~ 22 dB) is much higher than for entanglement distribution (in \ref{Result_Fig3}(d)) due to the inclusion of uplink turbulence loss. Equations used for calculation of the uplink turbulence loss are described in Section \ref{result_turb}.

% Optimum similar to L_0 proportional to d^2

\subsection{Simulation method}\label{simulation_method}

Light propagation through different optical elements used in our protocols is simulated using a python module named Lightpipe \cite{noauthor_lightpipes_2017-1}. We employed Lightpipe functions which uses fast Fourier transform (FFT) to numerically evaluate the propagated field using Fresnel approximation.

The electric field distribution $U(x,y,z)$ is related to the field angular spectrum $A(\alpha,\beta,z=0)$  through Fourier transform.

\begin{equation}\label{Az_Uz}
    A(\alpha,\beta,z) = \int \int_{-\infty}^{\infty} U(x,y,z)e^{ -ikz(\alpha x + \beta y)} dxdy
\end{equation}

In the Fresnel approximation, the propagated angular spectrum $A(\alpha,\beta,z)$ is related to the initial angular spectrum $A(\alpha,\beta,z=0)$ by the following relation \cite{goodman_introduction_2005},

\begin{equation}\label{H_Az_A0}
    H = \frac{A(\alpha,\beta,z)}{A(\alpha,\beta,z=0)} = e^{-i k z \sqrt{1-\alpha^{2}-\beta^{2}}}.
\end{equation}

This algorithm implements light propagation by calculating the angular spectrum $A(\alpha,\beta,z=0)$, given the field distribution $U(x,y,0)$ at z = 0, using FFT as  

\begin{equation}\label{A0_U0}
    A(\alpha,\beta,z=0) = \int \int_{-\infty}^{\infty} U(x,y,0)e^{ -ikz(\alpha x + \beta y)} dxdy.
\end{equation}

Once, $A(\alpha,\beta,z=0)$ is calulcated using Eq. (\ref{A0_U0}), $A(\alpha,\beta,z)$ can be easily found using Eq. (\ref{H_Az_A0}). The algorithm then calculates $U(x,y,z)$ from $A(\alpha,\beta,z)$ by using inverse Fourier transform (see Eq. (\ref{Az_Uz})), again employing FFT.

The algorithm implements FFT on a finite grid with periodic boundary conditions. Beam propagation using this method mimics light propagation in a waveguide the size of the grid \cite{noauthor_lightpipes_2017}. Due to the periodic boundary conditions used, if the field extends near the edge of the grid during propagation it is reflected. Hence, the algorithm is carefully implemented with large enough grid size that the field never propagates near the edge of the grid. 

In our simulation, after every beam propagation iteration either an aperture or a thin lens is used. A thin lens is implemented by an aperture (due to the lens finite size ) along with multiplying the field with the thin lens phase shift given by the formula,

\begin{equation}
    U'(x,y,0) = U(x,y,0)e^{ -ik\frac{(x-x_0)^2+(y-y_0)^2)}{2f}}.
\end{equation}

To calculate total loss in light propagation in Fig. \ref{Result_Fig3}, after considering the mentioned satellite loss for each satellite we also considered the following additional loss 

\begin{equation}\label{total_loss_eta}
    \eta  = \eta_{0}(1-e^{-\frac{-d^2}{2w_{LT}^2}}),
\end{equation}

where $d$ is diameter of the receiving telescope and $w_{LT}$ is the long-term beam width.
$\eta_{0}$ here consists of three important factors - $\eta_{e}$ which is transmission probability in presence of error calculated in Section \ref{error_appen}, $\eta_{a}$ which is atmospheric absorption and $\eta_{d}$ contains other losses like detector efficiency. They are related by the following expression:
\begin{equation}
\eta_{0} = (\eta_{e}^2) (\eta_{a})^2\eta_{d}    
\end{equation}

Hence, the above additional loss factor contains the uplink loss due to turbulence, atmospheric absorption loss, effect of setup errors and other losses at source and detector. The $\eta_0$ factor produced an additional 10 dB of loss which is the constant overhead loss for ASQN as it can be clearly seen in Fig. \ref{Result_Fig3}(d).

\subsection{Sources}\label{result_sources}

Along with signal loss, the other important factors in determining quantum communication rate are source, detector and electronics rates. In this respct, quantum communication rate or more specifically QKD clock rate, differs for qubit transmission using weak coherent pulses (WCP) and entanglement distribution due to different kinds of sources used. We are mentioning rates here without considering loss, i.e. rates at small communication distances. Loss is the principal rate limiting factor at large distances and this paper addresses that very issue through space transmission.

Entanglement distribution protocols generally use probabilistic entangled photon-pair generation like spontaneous parametric down conversion (SPDC) sources. Quantum dot based deterministic pair sources are emerging \cite{chen_highly-efficient_2018}, although they are still under development. Micius satellite used a SPDC source which can produce entangled photon pairs at a rate of 5.9 MHz \cite{yin_satellite-based_2017}. However, since then multiple new experiments has been performed \cite{zhao_high_2020, neumann_experimental_2022, meyer-scott_high-performance_2018} which shows much higher rates. Some of these sources employ frequency multiplexing capabilities too and achieve very high QKD secure key rate (more than 1Gbit/s) \cite{neumann_experimental_2022}. 

In qubit transmission either single photons or weak coherent pulses can be used to perform QKD. One of the recent experiments, which explores a device made by coupling quantum dots to photonic nanostructures, described a measurement of 10.4 MHz peak single photon detection rate from a generation rate of 122 MHz near-indistinguishable photons \cite{uppu_scalable_nodate}. Although, ideally single photons are needed for QKD, weak coherent pulses can be also used if decoy state \cite{lo_decoy_2005} QKD protocols are performed to prevent photon number splitting attacks. In Micius satellite a combination of four signal and four decoy states have been used for single photon generation at 100 MHz \cite{liao_satellite--ground_2017}. QKD clock rates as high as 2.5 GHz was achieved in optical fibers \cite{boaron_secure_2018}. A detailed review of quantum key distribution, including clock rates and secure key rates, can be found in \cite{xu_secure_2020}.

As discussed before in Section \ref{scheme},  qubit transmission and entanglement distribution protocols have their own unique advantages and challenges. The different forms of sources used in the two protocols and the different capabilities of the sources discussed above adds to these differences. The choice of sources is another thing to keep in mind while comparing between the two protocols.

Quantum satellite experiments mostly used polarization encoded qubits as photons' polarization states do not decohere while passing through the atmosphere \cite{liao_satellite--ground_2017,yin_entanglement-based_2020, toyoshima_polarization_2009}. However polarization qubits does decohere due to highly oblique reflection from spherical mirrors \cite{bonato_influence_2007}. This effect is pronounced in ASQN because of numerous reflections, especially when off-axis telescopes are used. Although, on-axis telescopes maybe more suitable, either validating or completely rejecting polarization qubits needs a much more detailed analysis which is out of the scope of the current work.

Hence, we explored alternative qubit designs for ASQN. Possibly the most interesting ones are time-bin and frequency-bin in qubits. Time-bin qubits should not be affected by either atmospheric turbulence or reflections from the moving satellites if the two bins are separated in time closely enough. For time separations shorter than other relavant times scales both bins would be affected identically keeping the qubit intact. For example, changes in atmospheric turbulence occur around every 10-100 millisecond \cite{cox_structured_2021}. Hence, time-bin qubits with bin separation in milliseconds or shorter should not be affected. Time-bin qubits \cite{brendel_pulsed_1999} have been used successfully for quantum communication quite often \cite{boaron_secure_2018} and even analyzed for satellite transmission recently \cite{jin_genuine_2019,vallone_interference_2016, bulla_non-local_2022}. The same argument holds for frequency bin qubits too. Here, the separation between the two frequency bins need to be small so that there is no significant change in the refractive index or reflectance values.

\subsection{Turbulence}\label{result_turb}
Atmospheric turbulence has important effects in quantum communication, especially for uplink transmission. The contribution to turbulence comes from the 20 km of atmosphere nearest to earth, with most contributions from near the surface \cite{bonato_feasibility_2009, fante_electromagnetic_1975, fante_electromagnetic_1980, pugh_adaptive_2020}. Consequently, in the downlink the beam doesn't have to propagate at all after the turbulent atmosphere resulting in negligible effect of turbulence loss compared to the diffraction loss \cite{rarity_ground_2002, sharma_analysis_2019}. In view of this, turbulence loss is generally not considered for downlink transmission. This would still be an approximation though, especially for downlink optimised beam profiles which are not necessarily flat beams. In special cases, there may be small effects of downlink turbulence. But as argued this contribution would not be anything significantly large. Considering this we didn't consider downlink turbulence loss. But diffraction loss in downlink is appropriately calculated and included as ground link loss.

% It would also be rather complicated to introduce downlink turbulence in our case as we have performed an optimisation for the diffraction loss in the ground.

In uplink transmission however the effect of loss due to turbulence is quite significant as the distorted beam emanating from the turbulent atmosphere has to propagate a long distance before reaching the receiver telescope on the satellite \cite{pugh_adaptive_2020}. Hence, the distorted beam gets very broad and the beam gets fragmented when it reaches the satellite \cite{villasenor_enhanced_2021}. The size of beam spot at satellite can be quantified \cite{bonato_feasibility_2009} by the long-term beam waist $W_{LT}$, with
\begin{equation}
    W_{LT}^2 = w_0^{2}\Bigg[1+\frac{L^2}{z_0^{2}}\Bigg] + 2\Bigg[\frac{4L}{k r_{0}}\Bigg]^2,
\end{equation}  
where $w_0$ is the beam waist emanating from the ground telescope, L = satellite elevation in meters, $k = 2\pi/ \lambda$ and $z_0 = \pi (w_0^2/\lambda)$ with light wavelength $\lambda$ and $r_0$ as Fried parameter or coherence length. The Fried parameter $r_0$ can be calculated using,
\begin{equation}
    r_0 = \Bigg[ 0.42 k^2 \int^L_0 C_n^2(z)\Bigg(\frac{l-z}{L}\Bigg)^{5/3} dz \Bigg]^{-3/5}
\end{equation} 

where 

\begin{multline}
    C_n^2(z) = 0.00594(v/27)^2 z^{10} 10^{-50} e^{-z/1000} + \\
    2.7 \times 10^{-16}e^{-z/1500} + Ae^{-z/100}
\end{multline}
    
is the atmospheric structure constant with A = 1.7$\times 10{-14}$ and v = 21 m/s \cite{bonato_feasibility_2009}. Upon calculating $W_{LT}$, the uplink loss can be calculated from Eq. (\ref{total_loss_eta}). $W_{LT}$ is almost independent of $w_0$ - as long as it is above a certain threshold - as the turbulence term ($2\Big[\frac{4L}{k r_{0}}\Big]^2$) dominates over the diffraction term in uplink. Hence, the choice of $w_0$ is not very significant \cite{pugh_adaptive_2020}. We have taken $w_0 = 25 cm$.

We incorporated the uplink turbulence loss in Fig. \ref{Result_Fig3}(e) with the changing satellite telescope diameter $d$, satellite to ground distance $L_{sg} = 200$ km, $\lambda = 800$ nm. Uplink tubulence loss is less when low elevation satellites (i.e., small $L_{sg}$) and large diameter telescopes (i.e. large $d$) are used. Both features come naturally in ASQN.

The above loss calculations are without considering any adaptive optics corrections, which can compensate for turbulence losses at least partially. Such adaptive optics corrections would be somewhat beneficial although at a cost of much more sophisticated systems like segmented mirror telescopes and laser guide stars \cite{pugh_adaptive_2020}. In adaptive optics corrections, $w_0$  and hence ground telescope diameters matters as bigger telescopes decreases the diffraction loss.

\begin{figure}
    \begin{subfigure}{}
      \centering
      % include first image
      \includegraphics[width=1.0\linewidth]{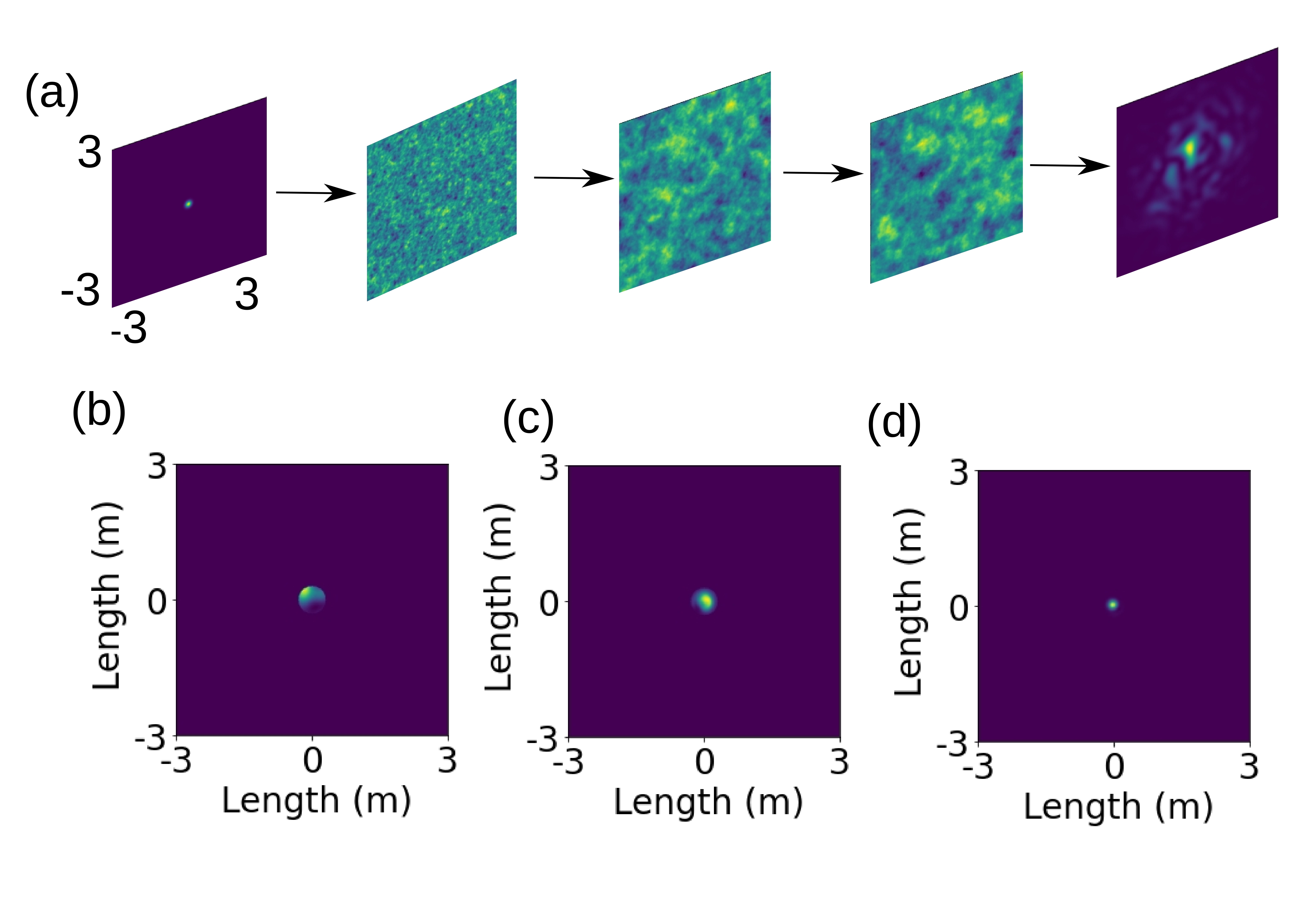}  
    %   \caption{Put your sub-caption here}
      \label{fig:sub-first}
    \end{subfigure}

    \begin{subfigure}{}
      \centering
      % include second image
      \includegraphics[width=.9\linewidth]{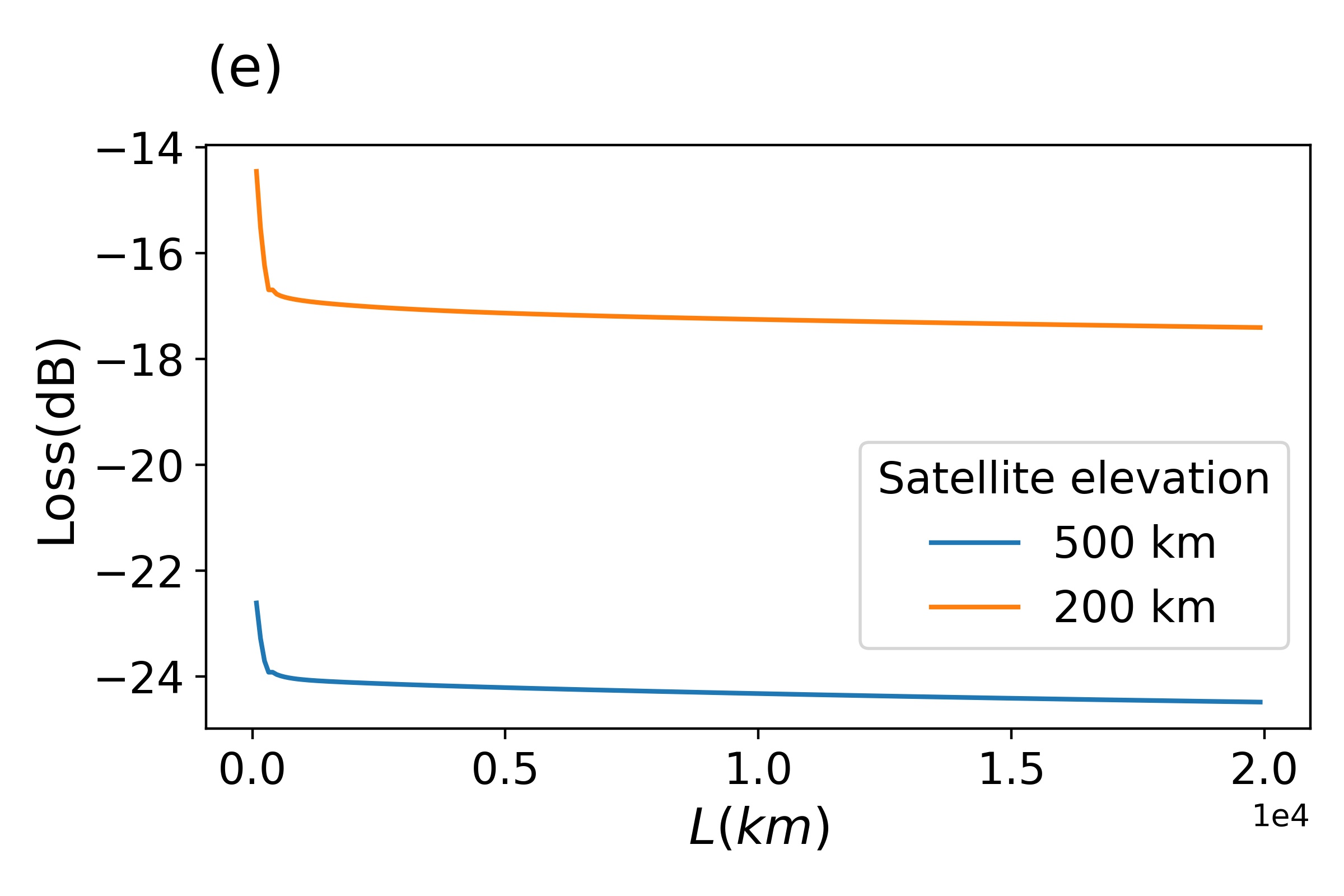}  
    %   \caption{Put your sub-caption here}
      \label{fig:sub-second}
    \end{subfigure}
    \caption{Effect of uplink turbulence in the qubit transmission proposal is numerically simulated. (a) Atmospheric turbulence, modeled using successive phase screens, completely fragments the initial Gaussian beam by the time it reaches the satellite. (b-d) The receiving telescope in the satellite only accepts a small part of the large fragmented beam (b) which is focused at a later satellite (c). This focused beam (c) is then transmitted through successive satellites (or ‘satellite lenses’) to as far as 20,000 km and the final beam (d) emerges largely unaffected from this long propagation due to effective confinement by the lenses. The light beam only suffers a small loss, much smaller compared to the initial turbulence-induced loss. (e) This is evident when average light propagation loss is plotted with propagation distance, over 300 iterations (with $d$ = 60 cm, $L_0$ = 80 km and for two cases of satellite chain elevation - 200 km and 500 km ).
}
    \label{turbulence_figure}
\end{figure}

Due to the effect of atmospheric turbulence a laser beam does not only get spread, the beam wavefront gets fragmented too \cite{villasenor_enhanced_2021, fante_electromagnetic_1975}. These effects may be seen as detrimental for further transmission of the beam through the satellite chain. However, we show through detailed numerical modelling that even in the presence of uplink turbulence and fragmentation of the beam, qubit transmission scheme (described in Fig. \ref{fig_scheme}(a) and Fig. \ref{Result_Fig2}(a) ) works quite well. Atmospheric turbulence is simulated using phase screens constructed following Kolmogorov's theory \cite{kolmogorov_local_1991}, implemented by the python module AOtools \cite{townson_aotools_2019}. These phase screens model the refractive index change and the corresponding phase imprint on the beam due to turbulence. The effect of turbulence is modelled using 17 phase screens, separated by certain specific distances following \cite{villasenor_enhanced_2021}, up to 20 km. The turbulence modelling process is described in Fig. \ref{turbulence_figure}(a) where the initial beam passes through the phase screens. Consequently the highly fragmented and spread out final beam profile is created.

However, the fragmented beam didn't affect the qubit transmission proposal substantially. In qubit transmission scheme described in Section \ref{result_simulation}, part of the uplink transmitted beam captured through the aperture is considered a constant wavefront. Although after including turbulence, this is not true anymore Fig. \ref{turbulence_figure}(b), the scheme described in Section \ref{result_simulation} still works effectively as the fragmented beam is focused generating a gaussian like shaped beam Fig. \ref{turbulence_figure}(b). The beam is not quite a Gaussian though as the imprints of turbulence still exist in the focused beam. However, ASQN is quite flexible with focal lengths, especially when large apertures are used. Please refer to focal length error discussion in Section \ref{error_appen}. Due to this focal length flexibility, even a distorted beam can be faithfully transmitted through the lens system over global distances. Final beam profile over propagation through 20,000 km (around 240 lenses with lens separation 80 km) is simulated and shown in Fig. \ref{turbulence_figure}(d). The beam stays almost intact at the center although its intensity drops a little.

Simulations in Fig. \ref{turbulence_figure}(b)-(d) is carried for one particular case, i.e. one particular random set of phase screens modelling turbulence. One such set of phase screens simulate turbulence only at one point of time. To find out the average loss due to the effects of turbulence, we performed the simulation 300 times and took the average. This would provide a better estimation of the average propagation loss for transmission up to 20,000 km. Losses other than diffraction and turbulence losses are not considered here. Propagation loss for the two cases of satellite elevation, 200 km and 500 km, are shown in Fig. \ref{turbulence_figure}(e). Diameter ($d$) of 'satellite lenses' was 60 cm, lens separation $L_0$ is 80 km and wavelength of light used was 800 nm. Average loss graphs in Fig. \ref{turbulence_figure}(e) clearly show that bulk of the loss is due to the initial turbulence effect while afterwards there is only a small loss over the lateral 20,000 km propagation. Hence, numerical simulations conclusively support the qubit transmission scheme reasonings described in section \ref{result_simulation}. Despite the devastating effect of air turbulence, focusing of the beam creates a tight spot of light which can be confined by successive 'satellite lenses' almost indefinitely. 

The detrimental effect of beam fragmentation still exist though, even distinct from the loss in the initial transmission. Influence of beam fragmentation in further propagation manifests itself by constraining the lens diameter $d$ and lens separation $L_0$ relationship. In presence of turbulence, complete light confinement by the lens system can only be achieved by using larger diameter or smaller separation lenses than needed otherwise. For example, 60 cm diameter 'satellite lenses' needs to be separated by 80 km instead of 120 km separation used in Fig. \ref{Result_Fig2}(b) in Section \ref{result_simulation}  when the turbulence effect was not considered. This effect occurs due to the distorted beam profile. A distorted beam, containing higher order modes, has a shorter Rayleigh range compared to an ideal Gaussian beam \cite{luxon_waist_1984}.

\subsection{Discussion on satellite orbits}\label{satellite_orbits}

In our simulation in Section \ref{result_simulation}, we have used a satellite elevation of 200 km as an example. Other satellite orbits are definitely possible for ASQN. Lower the satellite elevation lower would be diffraction loss in the ground link but being too low would also mean shorter satellite lifetime due to residual atmospheric drag in these orbits. Hence, to sustain orbit, continuous thrust is needed to be provided to the satellite. Such technology has already been demonstrated. The Gravity Field and Steady-State Ocean Circulation Explorer (GOCE) satellite \cite{noauthor_GOCE} sustained orbit in the 250 to 300 km range for 3 years using an ion propulsion system and the Super Low Altitude Test Satellite (SLATS) or Tsubame satellite \cite{noauthor_slats} operated in different orbits up to even 167 km elevations. The continuous thrust technology is still new though and the lower orbits are not widely adopted. This new technology may also have unknown implications for ASQN. In comparison, satellites can consistently maintain a stable orbit for years without additional thrust in higher orbits of around 500 km (still in low elevation in LEO which is around 200-2000 km). To see the effect of diffraction loss in such orbits due to the longer ground link, we have carried out another simulation at a satellite elevation of 500 km.

\begin{figure}
\begin{subfigure}{}
  \centering
  % include first image
  \includegraphics[width=.9\linewidth]{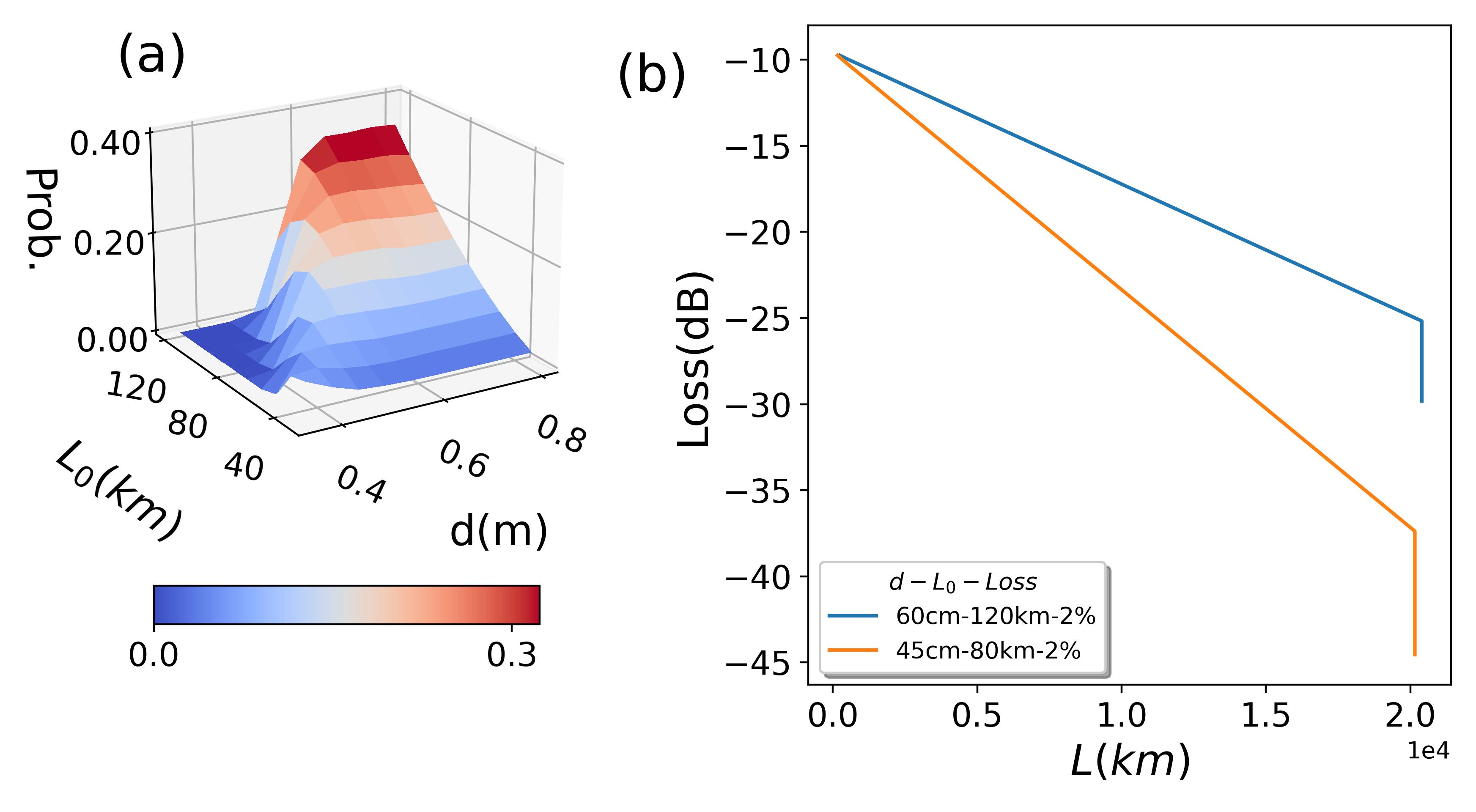}  \label{fig:sub-first}
\end{subfigure}
\begin{subfigure}{}
  \centering
  % include second image
  \includegraphics[width=.9\linewidth]{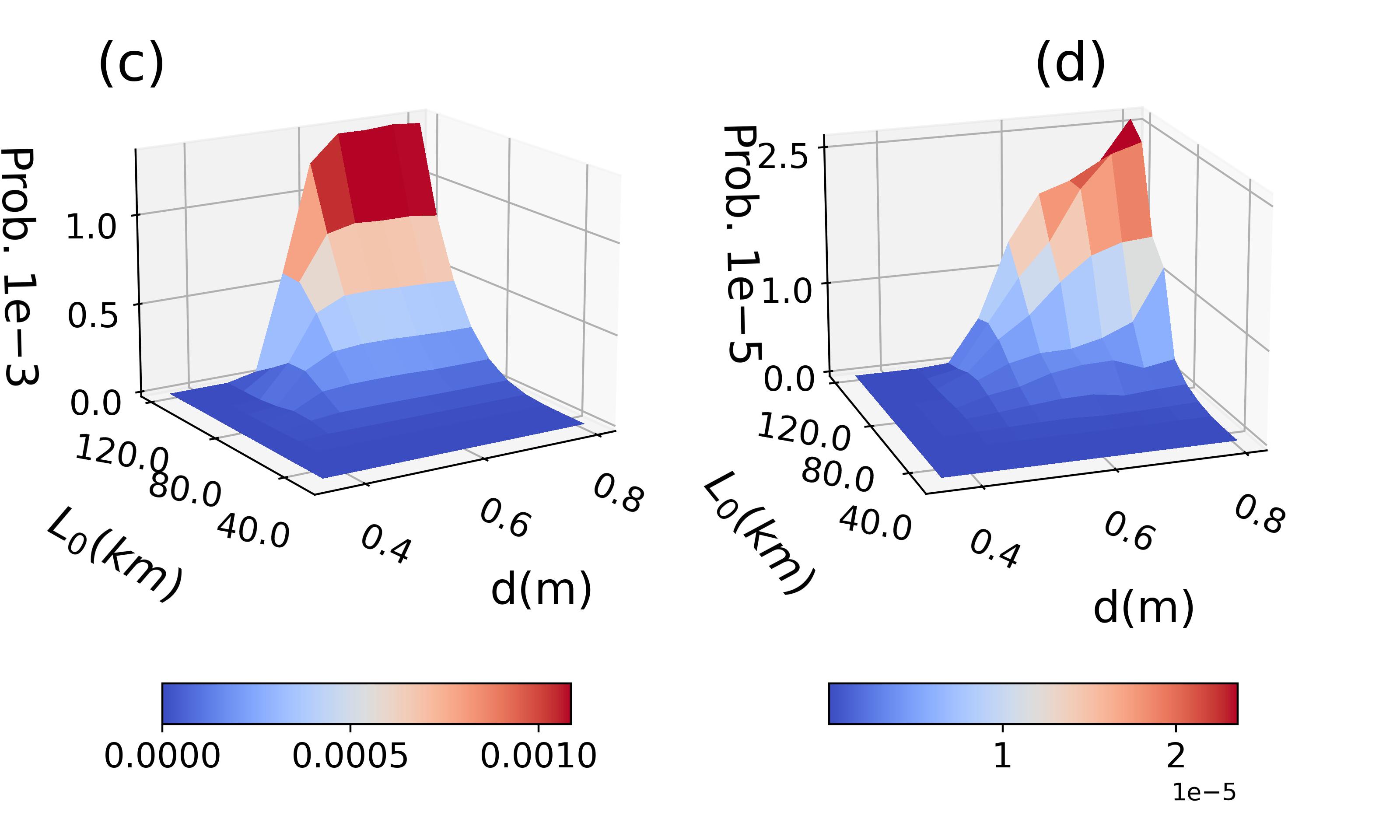}  
%   \caption{Put your sub-caption here}
  \label{fig:sub-second}
\end{subfigure}
    \caption{Simulation of light propagation in ASQN for 500 km satellite elevation - (a) Light transmission probability at 20,000 km in entanglement distribution protocol for different satellite telescope diameters ($d$) and satellite separations ($L_0$) are shown for ground telescope diameter of 1.2 m. (b)-(c) Total loss for entanglement distribution with 2$\%$ satellite loss showing an extra loss of only 3 dB compared to 200 km orbit. (d) The total loss for qubit transmission with 2$\%$ satellite loss increases a bit more (around 10 dB) due to the effect of uplink turbulence.}
    \label{Result_Fig500km}
\end{figure}

All the simulation parameters remain the same except satellite ground distance of 500 km and the ground telescope diameter of 1.2 m. Previously for 200 km elevation, the loss was so low that a smaller 60 cm diameter ground telescope was sufficient. However, a 1.2 m diameter telescope is quite common for ground operations and is the smaller of the 1.2 m and 1.8 m diameter ground telescopes used in the Micius quantum satellite experiments \cite{yin_satellite-based_2017}.

Diffraction loss in entanglement distribution only increased a little (see Fig. \ref{Result_Fig500km}(a)-(c)) as a much larger diameter ground telescope is used. For example, with 60 cm diameter satellite telescope and 120 km satellite separation ASQN diffraction loss increased by around 3 dB only (compared to the 200 km elevation case previously studied). This is achieved even without any focal length optimization, i.e. simply using fixed focal length values similar to Fig. \ref{Result_Fig1}(d). The loss contributes to a rather small change in the total loss (See Fig. \ref{Result_Fig500km}(b)-(c)). Entanglement distribution loss at 20,000 km changed from around 26 dB for 200 km elevation (as in Fig. 5 (b) and Fig. 5(d)) to around 29 DB for 500 km elevation (i.e., still less than 30 dB for $d$ = 60 cm, $L_0$ = 120 km) for 2$\%$ loss at each satellite. In the qubit transmission case, Fig. \ref{Result_Fig500km}(d) shows that about 10 dB extra loss is added (compared to the 200 km case in Fig. \ref{Result_Fig3}(e)) due to the effect of turbulence in the uplink (as already seen in Fig. \ref{turbulence_figure}(e)). However, as described in Section \ref{result_turb} we haven't considered any form of adaptive optics correction \cite{pugh_adaptive_2020} here which can definitely help in reducing the loss further. The effects of adaptive optics correction are naturally larger for larger elevations and can be around 8 dB\cite{pugh_adaptive_2020}. Also, in all cases of ASQN reflection loss is still the major factor which can be drastically reduced if ultra-high reflectivity Bragg mirrors can be used in telescopes.

In ASQN, ground links are vertical or near vertical as the long-distance light propagation happens through the satellite chain unlike for single quantum satellite experiments. For single satellites, highly oblique transmission with almost grazing incidence through the atmosphere is needed to distribute entanglement over long distances \cite{yin_satellite-based_2017}. In ASQN with 500 km satellite elevation even if one wants to communicate with a 500 km diameter circular area on the ground with the ground-pointing telescope that would mean sending light over a maximum of $\sqrt{500^2 + 250^2}$ km $\approx$ 560 km distance, only 12$\%$ more from the original 500 km vertical distance.

Different satellite orbits provide different sets of advantages and challenges. Lower orbits like 200 km elevation provide lower diffraction losses and hence better rates and also low satellite launch costs for the satellite chain as it needs to be elevated to a smaller height. These advantages are not present in higher orbits. Lower orbits also provide better space debris management. However, a 200 km orbit doesn't have stability without continuous thrust which a 500 km or even higher orbit would enjoy for multiple years. Also from higher orbits more of the ground is seen and can be communicated to near vertically which can be of importance especially transverse to the satellite chain. Hence, the choice of the orbit is not completely evident and would be more of an optimization between several factors.

\section{Influence of different factors}\label{Discussion}

\subsection{Mirror reflectivity}\label{mirrors}

To successfully carry out our protocol, one of the most important criteria is the minimal absorption loss during reflection from the system of telescopes. Absorption due to reflection scales exponentially with number of satellites and hence would cause huge total losses if absorption at each reflection can not be kept very small. We simulated total loss for 2$\%$ and 5$\%$ satellite absorption loss in Fig. \ref{Result_Fig3}. Each satellite would have a telescope system containing multiple mirror (as shown in Fig. \ref{Diss_Fig1}). Considering four mirror systems, each telescope mirror must have absorption loss less than 0.5$\%$ or 1.25$\%$ respectively (i.e., 99.5$\%$ or 98.75$\%$ reflectivity) to achieve total satellite loss below 2$\%$ or 5$\%$, if all four mirrors are from the same material. The best telescopes operating today generally do not need extremely high reflectivity though since they are mostly used for imaging purposes. For example, the gold coated beryllium reflectors of the James Webb telescope have 96.1$\%$ percent reflectivity for 800 nm light \cite{James_webb}. 

Among metals, bare gold and silver mirrors have high reflectivity. Gold and silver has respectively 98.4 $\%$ and 99.3 $\%$ reflectivity above 900 nm \cite{bennett_infrared_1965}. However, we must be cautious about the choice of reflective coating and the wavelength used for the beam.  Although for the above two metals, reflectivity increases with larger wavelengths, we cannot use light with arbitrarily large wavelength, for multiple reasons described below. The most important reason is as wavelength increases Rayleigh length ($z_R = \pi w_0^2/\lambda$) - which controls the beam diffraction - decreases. Hence, as $\lambda$ is increased, although reflectivity increases lens separation ($L_0 = z_R$) decreases requiring more lenses resulting in more loss. Atmosphere absorption is wavelength dependant too and hence constitutes of another constraint. In view of the above, an optimum working wavelength must be chosen carefully for specific reflective materials. 

Metal coatings are typically very delicate and require a protective coating. Reflectivity of protected gold and silver mirrors depends on the protective coating. In a broad wavelength range (700-2000 nm), with certain forms of protection gold has an average reflectance over 96 $\%$, whereas protected silver has over 98 $\%$ depending on two different forms of protection, in similar range \cite{noauthor_metallic_nodate}. This shows that the absorption loss in metal mirrors would be borderline for our purposes if only metal mirrors are used for the telescopes.

Another option would be to use Bragg mirrors, which achieve very high light reflectivity through the principle of constructive and destructive interference using alternating layers of two separate materials \cite{cole_tenfold_2013,chong_bragg_1992}. Bragg mirrors are considered to have the highest reflectivity in a range of wavelengths, with reflectivity as high as 99.9999$\%$ \cite{cole_high-performance_2016}. Although there are commercially available Bragg mirrors with 99.99$\%$ reflectivity \cite{paschotta_supermirrors_nodate}, they are generally of smaller radius ($\sim$ 10 cm) than required for our telescope back mirrors, with diameter of 40-60 cm. This implies there can be possible fabrication challenge to produce large Bragg mirrors. However, the smaller Bragg mirrors can be used as front mirrors along with larger metal back mirrors (as shown in Fig. \ref{Diss_Fig1}) and the reflectivity requirement of metal mirrors would be less stringent.

One place, where mirrors with such high reflectivity and such big size has been used is in Laser Interferometer Gravitational-Wave Observatory (LIGO) \cite{abbott2016observation, rakhmanov_dynamics_1998}. LIGO mirrors have reflectivity greater than 99.99$\%$ \cite{rakhmanov_dynamics_1998}. However, LIGO mirrors have much bigger constraints of being noise free \cite{harry_thermal_2002}, to carry out such sensitive experiments using their highly intense laser pulses. We do not have any such constraints as we are merely sending single photons and no ultra-sensitive measurements are required. ASQN only requires moderately large, high reflectivity telescopes to ensure the exponentially scaling absorption loss remains low over the complete satellite chain. For example, if 99.99$\%$ reflective mirrors are used in each of the four mirrors in as many as 200 satellites, a mere 8$\%$ light intensity will be lost to due to mirror absorption. Hence, just like diffraction loss, mirror reflectivity loss can be completely eliminated in ASQN too, even at global distances of 20,000 km.

The reflectivity values of Bragg mirrors for a particular wavelength hold for only a narrow angle of incidence and there is an issue of a huge spectral shift of reflectivity with large angle of incidence. Hence, the angle of incidence must be kept very low (ideally below 10$^{\circ}$) to achieve high enough reflectivity. A similar constrain on the angle of incidence arises due to polarization aberration. This is discussed in detail in Section \ref{new_aberr}. There are some broadband Bragg mirrors available though. Bragg mirrors with 10 cm diameter are commercially available and they have excellent reflectivity ($>$ 99$\%$)  for four different spectral ranges \cite{noauthor_fused_nodate}.
 
Regarding the choice of the shape of the mirrors, we must use continuous mirrors in all the telescopes, since segmented mirrors like the James Webb telescope \cite{James_webb} may give rise to large intensity loss and diffraction effects. That is one of the reasons why very large mirrors may be unrealistic for ASQN.

Since our protocol heavily relies on a set up consisting of multiple mirrors on multiple satellites, the damage and degradation of the mirror materials and the reflecting coating in space is a concerning issue \cite{sandin_materials_2021}. We mention in our protocol that we only need the satellites to be in Low Earth Orbits (LEO), even for a propagation distance up to 20,000 km. In Low Earth Orbits (LEO) at altitudes up to 700 km, atomic oxygen is the primary source of contamination for mirror materials. Due to the orbital velocity of satellites being around 8 km/sec, the material is exposed to atomic oxygens of around 5 eV energy. Although the application of thin-film protective coatings made of dielectric materials can reduce atomic oxygen related damage, a thorough study is necessary before choosing any mirror materials along with a protective coating \cite{garoli}.

Bragg mirror on the other hand doesn’t seem to be affected a lot by radiation as they are made of glass layers. There is generally only a certain amount of wavelength shift due to radiation. However, this conclusion is based radiation effects on Fiber Bragg gratings which are Bragg mirrors embedded in fiber \cite{gusarov_fiber_2000}. To the best of our knowledge, radiation effects on Bragg mirrors has not been thoroughly studied.

\subsection{Alignment, tracking and beam deviation}\label{tracking}

In ASQN reflector satellites move in a chain in the same orbit. Hence, they are co-moving and stationary with respect to each other. As they are stationary, these satellites in principle only need to be aligned instead of requiring dynamic tracking for light propagation. This is a major advantage of ASQN as dynamic tracking over such a large number of rapidly moving satellites would have made the protocol infeasible due to significant beam deviation (or tracking) losses at each satellite. However, there will still be some relative motion between the satellites due to them not being in the exact same orbits perfectly. Tracking may be needed, although it would be much slower and have much less stringent requirements. Hence, Beam deviation losses in such a system would primarily be due to satellite and telescope alignment losses. We quantified some aspects of such beam deviation loss in Section \ref{error_appen}.
 
Although tracking for all the individual satellites is not necessary for our protocol, we would need tracking in two specific cases, which are tracking towards the ground stations and for sending photons towards a satellite in a different chain, i.e., for a direction change of light needed for a 2D network of satellites. For ground tracking in the case of Micius, fine tracking accuracy of 0.4 $\mu$rad has already been achieved \cite{yin_satellite-based_2017}. Another reason 2-D network of satellites would be more complicated as satellite-to-satellite distance can change over time which can cause some transmission loss. It can be compensated though by dynamically adjusting the focal lengths of the mirror used in the satellite. Using this kind of network of 2D satellites a complete global quantum communication protocol can be achieved. However, even 1D transmission to global distances using one chain of satellites can be considered a very substantial achievement by itself.

For tracking and alignment, a high precision acquiring, pointing, and tracking (APT) system \cite{bourgoin_comprehensive_2013} is needed. In ASQN the tracking beam must pass through the same path as the transmitted photons, going through multiple satellites from the source to collection points. There is no other way to do the tracking with only one tracking beam.  Dichroic mirrors, which reflects at a particular frequency and transmits at the others, can be used for this purpose. These dichroic mirrors can be dynamically brought in or out of light path as needed. For example, all the dichroic mirrors can be brought in light path initially to align the satellite chain before starting any signal transmission. Later, when signal transmission starts only some of them can be left in light path (in a specific satellites) for dynamic tracking. Progress in artificial intelligence may also be useful for handling complicated tracking and alignment issues quickly. Instead of dichroic mirrors, we can alternatively use beam splitters with very small reflectivity (say less than 0.1 $\%$) that would cause very small photon loss for the signal. However, for alignment large power will be needed for the tracking beam.

\subsection{Telescope setups, vortex beam and focal length}\label{telescope_setups}

\begin{figure}
\begin{subfigure}{}
  \centering
  % include first image
  \includegraphics[width=.9\linewidth]{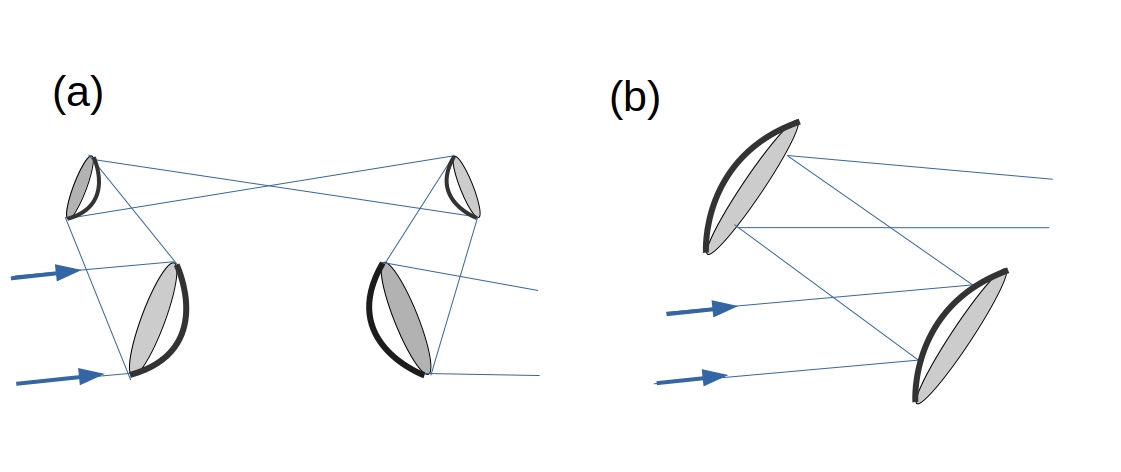}  
%   \caption{Put your sub-caption here}
  \label{fig:sub-first}
\end{subfigure}

\begin{subfigure}{}
  \centering
  % include second image
  \includegraphics[width=.9\linewidth]{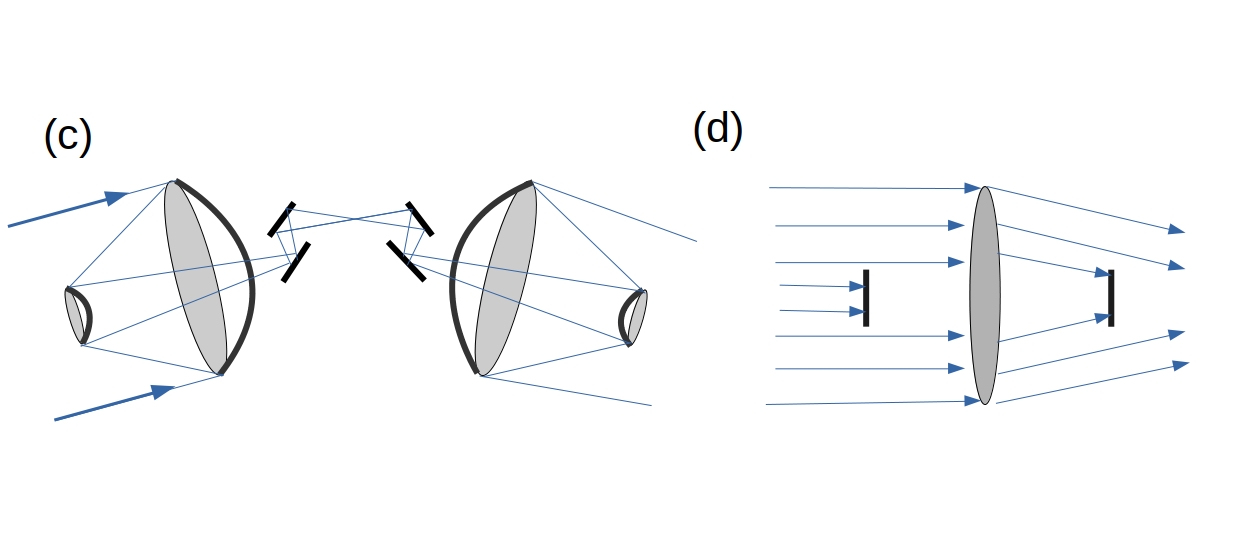}  
%   \caption{Put your sub-caption here}
  \label{fig:sub-second}
\end{subfigure}
\begin{subfigure}{}
  \centering
  % include second image
  \includegraphics[width=.9\linewidth]{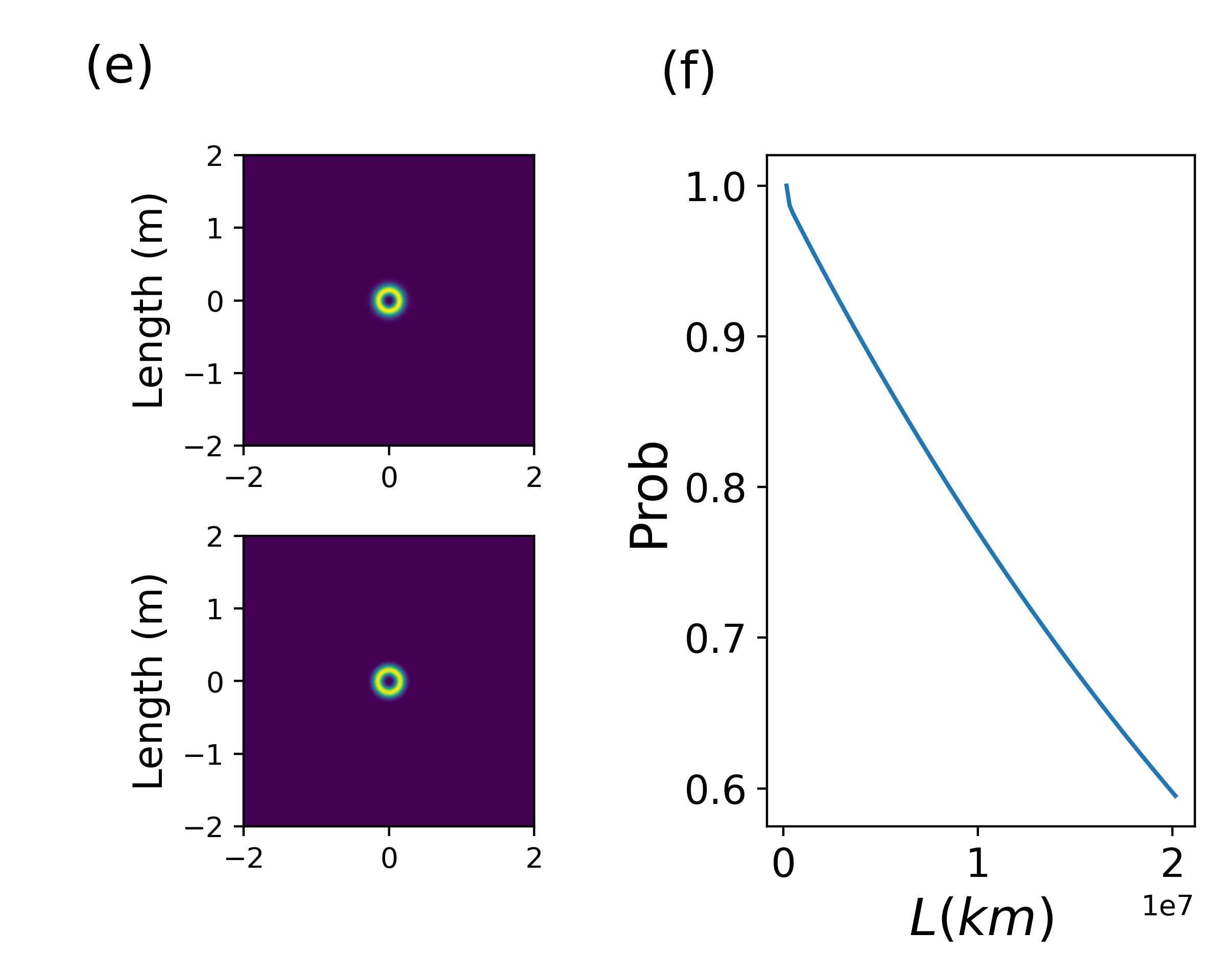}  
%   \caption{Put your sub-caption here}
  \label{fig:sub-second}
\end{subfigure}
% \begin{figure}[htbp]
%     \centering
%     \includegraphics[width=0.95\linewidth]{Result_Fig3.jpg}
    \caption{(a)-(d) The different possible telescope setups suitable for a chain of satellite reflectors are depicted, showing beam focusing and bending. (a) and (b) shows off-axis telescope setups while (c) shows an on-axis setup. The on-axis setup (without the folding mirrors) can be modelled by the screen-lens-screen system shown  in (d). Diffraction loss through such an on-axis system can be contained by using vortex beams. (e) A simulation of entangled photon pair propagation through 20,000 km in vortex beam profile through the on-axis system shows nearly identical initial and final beams (after 10,000 km propagation by each photon). (f) Entangled pair transmission probability is plotted with distance  (ground link or other losses not included).}
    \label{Diss_Fig1}
\end{figure}

Different forms of telescope setups possible in a reflector satellite are shown in Fig. \ref{Diss_Fig1}.  Fig. \ref{Diss_Fig1}(a) and (b) shows off-axis telescopes while Fig. \ref{Diss_Fig1}(c) shows an on-axis one. Fig. \ref{Diss_Fig1}(a) and (c) are four mirror setups with two telescopes each having a front and back mirror. These back mirrors have significant curvatures, enough to focus light at a short distance ($\sim$ 1 m). In contrast Fig. \ref{Diss_Fig1} (b) is only a two-mirror set up with mirrors only slightly curved (focal length around 70 m). Each of the three setups has their unique advantages and disadvantages which would be explored throughout this subsection. Off-axis telescopes, in general, face harsher beam aberration effect then their on-axis counterparts. But off-axis telescope setups have their own advantages. In ASQN, light needs to be bent slightly at each satellite to achieve light propagation along the curvature of Earth. The slight bending required at each satellite can be naturally accomplished in the off-axis setup, without requiring additional mirrors, as seen in Fig \ref{Diss_Fig1} (a) and (b). On-axis telescopes would however need additional plane mirrors (or fold mirrors) in the middle to provide the bending. Additional mirrors would cause more loss and polarization aberration, although these effects can be limited by using Bragg mirrors and small incidence angles. Another major advantage of off-axis telescopes have is that they don't obstruct incoming light beam as the front mirror in an on-axis telescope do. This reason makes off-axis telescope a much better fit for our protocol as successive obstruction causes photon loss. However, one can use an on-axis telescope too by using a vortex beam as shown in Fig. \ref{Diss_Fig1}(e)-(f) and explained below.
 
All the previously shown simulations assume off-axis telescope setups as no light obstruction due to front mirror is assumed. An on-axis telescope would obstruct the central portion of a Gaussian beam. To transmit the whole beam a vortex beam can be chosen instead. Vortex beams have doughnut
shaped intensity profiles (as shown in Fig. \ref{Diss_Fig1}(e)) and are represented by the following equation,
\begin{multline*}
    E(\rho,\phi,z) = E_{0} \frac{w_{0}}{w(z)}
    \exp{\Bigg[-\frac{\rho^{2}}{w^{2}(z)}\Bigg]}\\ \exp{\Bigg[-ikz-i\frac{k\rho^{2}}{2R(z)} + i\eta(z)}\Bigg] \exp{(im\phi)},
\end{multline*}
with vortex charge m \cite{allen_orbital_1992}. Here, $k = \frac{2\pi}{\lambda}$ is the wave number ,$w_{0}$ is the beam waist radius, $w(z) = w_{0}\sqrt{1+(\frac{z}{z_{R}})^2}$ is the spot size parameter, $E_{0}$ is the electric field amplitude at origin $(\rho=0,z=0)$, $R(z) = z\Bigg[1+(\frac{z_{R}}{z})^2\Bigg]$ is the radius of curvature at position z and $\eta(z) = \arctan(\frac{z}{z_{r}})$ is the Gouy phase while $z_{R} = \frac{\pi w_{0}^2}{\lambda}$ is the Rayleigh range for the beam.  

The On-axis setup in Fig. \ref{Diss_Fig1}(c) can be modelled by a lens put between two screens, as shown in Fig. \ref{Diss_Fig1}(d). The screens model the obstruction of the light by front mirrors while the lens models the effective focusing due to the four mirror surfaces, just like the off-axis setups. We ignored the bending of the light in our model, similar to off-axis models in Fig. \ref{Result_Fig1} and Fig. \ref{Result_Fig2}. Hence, the folding plane mirrors are excluded. A chain of such satellites with on-axis telescopes are modelled by a chain of screen-lens-screen setups (as shown in Fig. \ref{Diss_Fig1} (d)) through which beam propagation is simulated. Beam and telescope parameters can be adjusted, so that the vortex beam is transmitted perfectly. For telescope back mirror diameter ($d$) of 60 cm, satellite separation ($L_0$) of 80 km, vertex charge ($m$) of 2, and front mirror diameter of ($d/10$) 6 cm we have a near-perfect transmission of vortex beam with transmission probability of 0.6 at 20,000 km (without considering ground link). The initial and final vortex beam profiles are shown in Fig. \ref{Diss_Fig1}(e). Only diffraction loss is considered here. They have very slight differences, with the final one a little dimmer and a bit truncated in the inner edge. This confirms faithful propagation of the beam. In Fig. \ref{Diss_Fig1}(f) the decrease in transmission probability with distance follows a similar progression as Gaussian beams in Fig. \ref{Result_Fig2}(c). However, the beam parameter do differ as Rayleigh range is no more $z_R = \pi w_0^2/\lambda$ which holds only for an ideal Gaussian beam. For, higher order modes Rayleigh range decreases  \cite{luxon_waist_1984} and hence  satellite separation ($L_0$) needs to decrease too to contain diffraction loss. 

In modelling the vortex beam transmission through on-axis setup, we ignored the front mirror holders. If these mirror holders are present they would inevitably come in the light path (as opposed to the off-axis telescope case) and affect diffraction at each satellite. However, as the satellites would be weightless in LEO the front mirror holders are not absolutely indispensable. The front mirror holders can be detached during final transmission. Alignment of the front mirror can be adjusted physically using the holder, after they are detached but before completely withdrawing them. Otherwise one may try to do it remotely through electromagnetism. Also a small misalignment can be corrected by aligning the plane fold mirrors accordingly.

Another important consideration is how the telescopes in Fig. \ref{Diss_Fig1}(a)-(c). needs to be adjusted to provide the required focusing. The effective focal length created by the mirror setups can be adjusted by changing mirror positions. In the setups of Fig. \ref{Diss_Fig1}(a) and Fig. \ref{Diss_Fig1}(c), it is possible to adjust mirror positions to keep parallel rays parallel, due to the inherent reflection symmetry of the setups. Then these setups effectively behave as apertures (or lenses with infinite focal length). However, on shifting mirror positions focal lengths would change which is calculated below. The four mirror setups in Fig. \ref{Diss_Fig1}(a) and Fig. \ref{Diss_Fig1}(c) can be modelled by two lenses if the two front mirrors are considered as plane mirrors. If this is not assumed, the analysis would be a bit more complex but the result would be similar. Curved mirrors are equivalent to lenses. The two back mirrors are considered as two lenses with focal length $f_1$ and $f_2$ separated by a distance $d$. The compound or effective focal length of this lens system measured from the back mirror (with focal length $f_2$) is given by $f_e = \frac{f_2(d-f_1)}{d – (f_1+f_2)}$ \cite{hecht_optics_2002}. If parallel rays are kept parallel, $f_e = \inf$, implying $d = (f_1+f_2)$.  The two curved back mirrors (and hence the lenses) are assumed to have the same focal length $f_1 = f_2 = f $, $d = 2f$. If the separation is changed by $\Delta$, i.e. from $d$ to $d + \Delta$, the effective focal length becomes $f_e = \frac{f(f+\Delta)}{\Delta} = \frac{f^2}{\Delta}$ if $f >> \Delta$. In Fig. \ref{Diss_Fig1}(a) and Fig. \ref{Diss_Fig1}(c) separation between back mirrors can be increased by shifting front mirrors only. To achieve $f_e$ = 50 km with $f$ = 1 m one must have $\Delta = 20 \mu m$. Although it may look challenging to shift the front mirror ($\sim$ 10 cm diameter and 5 kg weight) by such a small amount, commercial piezo motors achieve this regularly on Earth \cite{lp_p-752_nodate}. However, even then there would be some errors in focal length. Fortunately, the effects of a \textit{f} error grow slowly with $10\%$ focal length error decreasing transmission probability only by 4.4 dB, as discussed in Section \ref{error_appen}.
 
We now discuss the case of Fig. \ref{Diss_Fig1}(b), where $f_1$ and $f_2$ have opposing signs as the mirrors have opposite curvatures. However, one can still achieve $ d = f_1 + f_2$ with $f_1$ and $f_2$ having slightly different values if $|f_1|, |f_2| >> d$. For small shifts $\Delta$ in separation $f_e = \frac{f_2(f_2+\Delta)}{\Delta} = \frac{f_2^2}{\Delta}$ if $f_2 >> \Delta$. To achieve $f_e$ = 50 km, one may set $f$ = 70 m and $\Delta$ = 1 cm. So, we can have near zero curvature mirrors while the mirror position adjustment would be easier.

\subsection{Aberration}\label{new_aberr}

The aberration of an optical system can be seen as the deviation from its ideal working scenario. Telescope mirrors shown in Fig. \ref{Diss_Fig1}(a)-(c) can introduce different forms of aberration like geometric or wavefront aberration, chromatic aberration, and polarization aberration. Wavefront aberration is caused by the variation of optical path lengths due to the geometry of the reflecting or refracting surfaces. Chromatic aberration originates due to frequency dependant refractive index or reflection coefficient of an optical element. Polarization aberration occurs because of the variation of two polarization components (s and p) in refracted or reflected light depending on the angle of incidence.
 
Due to wavefront aberrations from curved mirrors, a light beam in a certain spatial mode goes through mode transformations. This effect is especially pronounced when the light ray’s incident at an angle to the curved mirror axis, as it is by construction for an off-axis telescope. In case of an on-axis configuration of the optical system, overall aberration effects are much less pronounced.  The angle of acceptance (i.e., the maximum angle the light beam makes with the telescope axis) for 60 cm diameter telescopes separated by 120 km is (60cm/120km) = 1/20000 $\sim $ 1/1000$^{th}$ of a degree, for an on-axis setup. Therefore, the aberration effects would be correspondingly smaller.
 
Wavefront aberration consists of different forms like spherical aberration, coma or astigmatism. However, geometric aberrations being beam spatial mode transformations due to reflection these can be compensated by arranging reflecting surfaces properly, although it is very difficult to compensate all forms of aberration together. One such aberration compensating system is a three-mirror anastigmat \cite{korsch_closed_1972} which correct coma, astigmatism, and spherical aberration together to a large degree and can be employed in ASQN to. In this context, the James Webb telescope also uses a three-mirror anastigmat \cite{James_webb}.

Plane mirrors on the other hand produces no wavefront aberration. For this reason, plane mirrors are used as fold Mirrors used in Fig. \ref{Diss_Fig1}(c) are plane mirrors. Although, plane mirrors do not provide beam focusing either which is necessary for ASQN. But slightly curved mirrors with large focal length as in Fig. \ref{Diss_Fig1}(b) may be adequate for this task, that is producing focusing while reducing aberration. In a nutshell, the off-axis telescope setup with highly curved mirrors in Fig. \ref{Diss_Fig1}(a) is supposed to suffer the highest amount of geometric aberration while that of Fig. \ref{Diss_Fig1}(b) with only slightly curved mirrors should produce less aberration and the on-axis setup of Fig. \ref{Diss_Fig1}(c) should produce the least aberrations, especially if three-mirror anastigmat setup is used. Simulations in Section \ref{result_simulation} didn't consider any effects of aberration as the telescope systems are modeled as single lens. Even if the chain of satellites has a substantial aberration effect for the off-axis setup of Fig. \ref{Diss_Fig1} (a) there are several options to explore especially given the possible vortex beam propagation (shown in Fig. \ref{Diss_Fig1}(e)-(f) and discussed earlier) through the on-axis setup.
 
Chromatic aberration occurs in telescope setups due to variations of reflection co-efficient with light wavelength. All our simulations were carried out for monochromatic light. So, the effects of chromatic aberration were not studied which will manifest when considering a pulsed light. However, if the variation of reflection coefficient is small over the pulse bandwidth chromatic aberration can be neglected. Along with ignoring chromatic aberration, another approximation in considering monochromatic light for diffraction simulations is that pulsed light may have different diffraction effects compared to a monochromatic wave. However, several studies showed that pulsed light influences diffraction only when its frequency bandwidth reaches near light frequency i.e., for near femtosecond optical pulses \cite{gu_fresnel_1996}.

Polarization aberration is an important concern for our setup as aberration would occur at each satellite that would cause an overall large effect as explained below. If polarization qubit is used for encoding such aberration effects can potentially completely decohere the qubit \cite{bonato_influence_2007}. As discussed in Section \ref{result_sources} we can navigate this problem by using other form of qubits like time-bin or frequency-bin qubits. However, polarization aberration remains a concern even then as it may cause geometric aberration effects too which may affect diffraction.

Polarization aberration of a reflecting surface depends on two quantities, retardance and diattenuation which in turn depends on the reflection coefficients \cite{breckinridge_polarization_2015}. As both retardance and diattenuation varies quadratically with the angle of incidence, polarization aberration can be significantly diminished by reducing the largest angle of incidence. For highly reflective surfaces (say, metal or Bragg mirrors) polarization aberration is generally negligible below an angle incidence angle of 10 degrees \cite{bonato_influence_2007}. The largest angle of incidence on the telescope front and back mirror depends on the distance between the two mirrors, their sizes, and their curvatures. If one tries to achieve a small angle of incidence at too close a distance between the mirrors, the front mirror moves into the light path and creates obstruction. The telescope setup of Fig. \ref{Diss_Fig1}(a) can achieve a 10-degree largest angle of incidence if the separation between front and back mirrors is around 1 m (assuming a flat front mirror). The required distance would be even more for Fig. \ref{Diss_Fig1}(b) as both the mirrors are of large size. Hence, polarization aberration would be an important issue for this setup. In Fig. \ref{Diss_Fig1}(c) the flat fold mirrors do not need to have a lot of distance between them as they are of small size, although large enough for the focused light beam. The protective coating film on mirrors can be adjusted to reduce polarization aberration too \cite{breckinridge_polarization_2015}.
 
In conclusion, different telescope setups would have different advantages and challenges. The off-axis setup of Fig. \ref{Diss_Fig1}(a) would provide no obstruction, easy bending of light in two dimensions, small polarization aberration while there may be detrimental geometric aberration effects and fine mirror position adjustment would be needed for focusing. Mirrors in Fig. \ref{Diss_Fig1}(b) would possibly have smaller geometric aberration effects and crude mirror adjustments would be sufficient for focusing although there may be larger polarization aberration and other difficulties in using a two-mirror setup, e.g. difficulty in aligning for ground link transmission. The on-axis setup in Fig. \ref{Diss_Fig1}(c) would probably suffer the least geometric aberration but to bend the light multiple fold mirrors would be needed which would increase both reflection loss and polarization aberration. The obstruction of light by the front mirror would necessitate the use of vortex beam instead of Gaussian beams needed for earlier setups. Considering this myriad of benefits and challenges different telescope setups can be used in different satellites in the same chain. For example, Fig. \ref{Diss_Fig1}(a) telescope setups can be used in special satellites for ground links while Fig. \ref{Diss_Fig1}(b) setups in most of the satellites to control aberrations in propagation.

The above discussion was focused only on reducing aberration. However, we have not delved into the discussion of what happens to the diffraction loss if a large aberration effect is actually present in the beam (i.e., the beam is far from a ideal Gaussian shape and higher modes are present). It is quite possible that even significant aberration effects doesn't affect the diffraction loss a lot. One promising example in this direction is the propagation of the fragmented uplink turbulent beam shown in Fig. \ref{turbulence_figure} Although the fragmented beam was focused, its shape (in Fig. \ref{turbulence_figure}(c)) is far from an ideal Gaussian. Still the beam propagated to 20,000 km under the very same lens configuration with only a little less satellite separation distance ($L_0$). However, to truly ascertain the magnitude and effect of aberration one needs to perform further detailed theoretical analysis and possibly needs experimental confirmation too, which is discussed next.

\section{Possible Tabletop Experiments}\label{tabletop}

Along with theoretical analysis, influence of many of the factors discussed above can be verified using laboratory experiments which can be achieved by using mirror setups like in an actual satellite chain while scaling down the sizes in the setup of Fig. \ref{Result_Fig1}(a) - decreasing both $d$ and $L_0$. For example, if we want the Raleigh range to be restricted to a distance manageable in laboratory experiments (say, 1 m), a 800 nm laser beam need to have a beam waist of around 0.5 mm. Considering the diameter of mirrors (lens in Fig. \ref{Diss_Fig1}(a)) $d = 4 w_0$, mirrors of size 2 mm are needed. This would be the size of back mirrors (as shown in Fig. \ref{Diss_Fig1}) while front mirrors would be even smaller. Micro mirrors of diameter in mm or even much smaller are in use in different optical device applications \cite{douglass_lifetime_1998}. Ultra small parabolic mirrors have also been developed \cite{morozov_objective-free_2020}. Hence it should be possible to adapt this mirrors to our setup. Although, if necessary even larger scale ($\sim$ cm size) mirrors can be used by fitting an appropriate sized aperture right next to them although that may cause some additional diffraction effects depending on the fitting of the aperture to the mirror. One may also try this experiment using beams with larger beam waist ($\sim$ 1 cm) but then it can not be confined within the lab as the Rayleigh range would increase to hundreds of meters.

A miniaturize setup can verify many aspects of the real satellite chain while being much easier to develop and low-cost. These include effects of beam truncation, aberration and polarization aberration, beam deviation, mirror reflectivity loss (especially effects of the protective coating), comparing effectiveness of different telescope setups, focal length error, satellite position error etc. One may even investigate qubit transmission by introducing artificial turbulence on the beam by using special light modulator\cite{toselli_slm-based_2015}. Some things would however be different in lab like effect of gravity on the set up (weightless in orbit), errors due to satellite velocity (probably difficult to simulate in lab), adjustments required to accommodate large size mirrors (if used).

\section{Conclusion}\label{concl}

Inspired by rapid developments in space technologies, we presented a satellite-relayed space-based quantum communication protocol which would have low loss even at global distance of 20,000 km due to elimination of diffraction loss by the 'satellite lenses'. This architecture (ASQN) would be most important to establish quantum communication (and especially QKD) at large distances (5,000-20,000 km), although ASQN can be the least lossy protocol over almost the whole range of distances available (200 -20,000 km) due to the progressive increase in loss. This architecture does not require quantum memories whose development has been the major roadblock for quantum networks. Although challenging, launching such a chain of satellites seems plausible in view of the recent advances in space technology. In absence of diffraction loss, other losses (due to mirror reflectivity, beam deviation, aberration, turbulence etc.) become the important factors. These effects were investigated in significant detail and when applicable possible avenues are suggested to limit their detrimental effects. For example, in on-axis telescope setups Gaussian beams would be obstructed at the front mirror. So, vortex beam transmission was considered and successfully simulated. Further exploration in the form of both theoretical analysis and proposed laboratory experiments would be helpful
to confirm the influence of all factors on photon transmission. In addition to entanglement distribution, we also discussed 'qubit transmission' protocol which seems to have multiple advantages by having both sources and detector on ground. Even after considering the adverse effect of uplink turbulence, we show that 'qubit transmission' can deliver comparable or even larger rates than regular entanglement distribution.

\section*{Acknowledgement}

SG would like to thank Christoph Simon for teaching quantum networks.

\section{Appendix}\label{appendix}

\subsection{Quantum Internet}\label{QI_appen}

\begin{figure}
     \centering
     \begin{subfigure}%[b]{0.3\textwidth}
         \centering
         \includegraphics[width=0.5\textwidth]{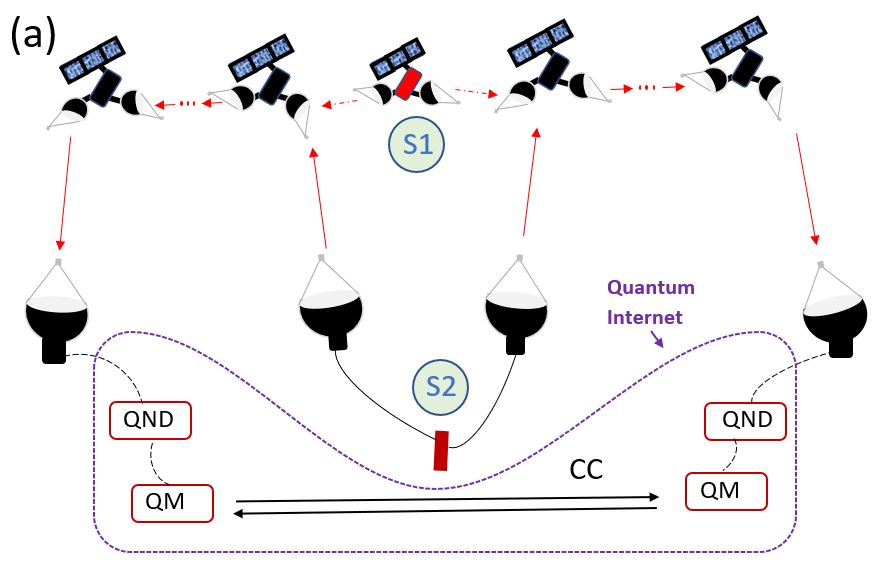}
         \label{fig:direc transm}
     \end{subfigure}
     \hfill
     \begin{subfigure}%[b]{0.3\textwidth}
         \centering
         \includegraphics[width=0.5\textwidth]{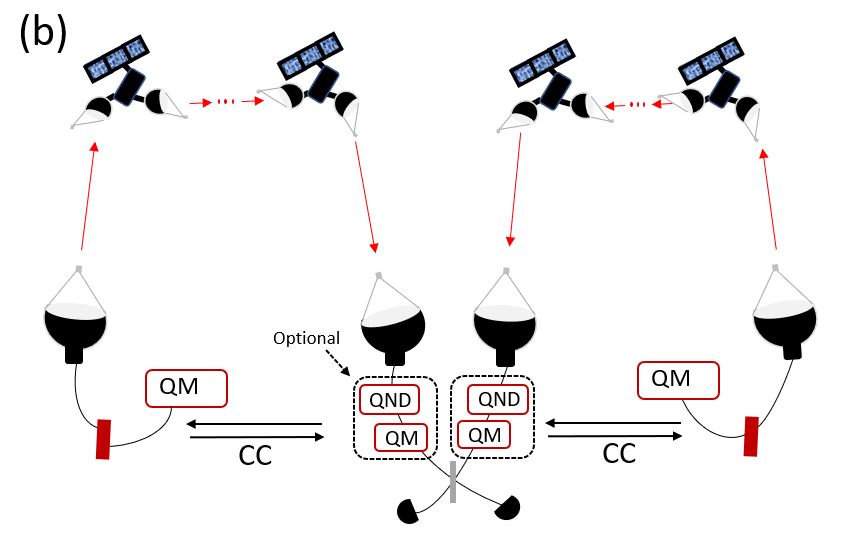}
         \label{fig:entangl_distr}
     \end{subfigure}
     \hfill
     \begin{subfigure}%[b]{0.3\textwidth}
         \centering
         \includegraphics[width=0.5\textwidth]{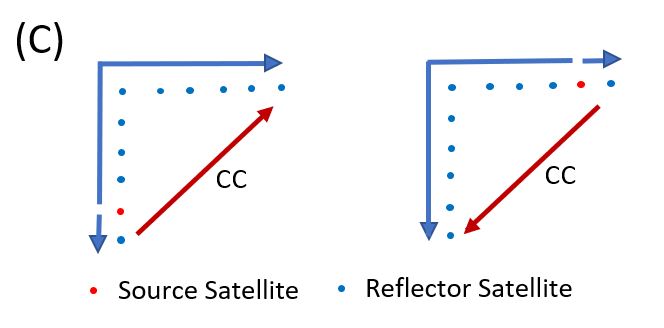}
         \label{fig:quantum internet}
     \end{subfigure}
\caption{Quantum internet protocol using ASQN – (a) Quantum internet by entanglement distribution with entangled source on satellite (S1) and on ground station (S2). Transmitted entangled photons can be collected in ground stations to establish entanglement deterministically for use in quantum internet schemes. This requires quantum memories (QM), quantum non-demolition (QND) measurements and classical communication (CC). See text for details. (b) Quantum internet using repeater scheme – Quantum internet can be established by doing a Bell state measurement using ground based detectors in the middle without needing QND detectors. However, the entanglement generation rate can be increased multifold by using the optional QND detector and quantum memory in the middle. (c) Multiple teleportations without using quantum memories. Using low-loss entanglement distribution protocol asymmetrically (with unequal arms) a qubit can be teleported to one place and teleported back (or sent to another place), with the help of redundancy.
}
\label{QI_Appendix_Fig}
\end{figure}

Quantum internet capabilities can be built using our protocol, although quantum memories would probably be needed. Some of the stringent requirements on quantum memories (when used in a repeater protocol) can however be relaxed due to elimination of diffraction loss. The basic building blocks of a quantum internet would be heralded entangled pairs distributed between two far away points. These entangle pairs can then be used for distributed quantum computing or quantum sensing applications.

Memories are required for two principal reasons. Firstly, memory is needed to store the qubit while one waits for confirmation on the heralded entangled pairs. For heralding, there can be two main strategies - either using sophisticated quantum non demolition (QND) detectors at each end (Fig. \ref{QI_Appendix_Fig}(a)) or detecting photons in the middle (Fig. \ref{QI_Appendix_Fig}(b)). In both cases, one would need to receive the detection results (either from the other end or from the middle) at each ground station to know which photons are really part of an entangled pair and hence can be used for quantum internet. The other reason is more fundamental and requires memories even for lossless transmissions which would not require heralding as no entangled pairs would be lost. In general quantum internet protocols would require quantum operations conditioned on classical communication of measurement results from the other end to obey causality (like in quantum teleportation). Hence, the optical qubits need to be stored during the classical communication. Another reason memories would be good for building quantum internet is that they would boost the rate like in repeater protocols, if used that way. This is optional as diffraction loss is eliminated already, although it can be important when uplink turbulence loss is included (as in the setup of Fig. \ref{QI_Appendix_Fig}(b))

Using memories, one strategy to create a quantum Internet with entanglement distributed from a source in the middle is shown in Fig. \ref{QI_Appendix_Fig}(a). Here, QND detectors are used to non-destructively measure which photons have successfully arrived and then these photons are stored in quantum memory. In QND detection photons are neither destroyed nor their quantum state is measured. Only the presence or absence of the photon is measured. With QND detection less multimode capacity would be required of the quantum memories as storage capacity would need to be arranged only for the arrived photons and not all photons including the lost ones. For long distance links high storage time ( $\sim$ 100 ms) quantum memories would be needed. High memory efficiency is not essential as there are only two memories involved (e.g., at 50$\%$ efficiency the rate will drop by a factor of 0.25, a loss of 6 dB which may not alter the total loss significantly), in contrast to the many memories needed in a repeater scheme.

The other strategy to make a functioning quantum Internet would be to create one link of quantum repeater using ASQN, as shown in Fig. \ref{QI_Appendix_Fig}(b). This would not require a QND detector but would require memories with high storage time and multimode capacity. Again, high efficiency would not be a necessity. In this architecture, photons will be sent from the two end stations and entanglement would be created and heralded by Bell state measurement (BSM) at mid-point. Different schemes employed in fiber-based quantum repeaters like Barret-Kok scheme, DLCZ scheme, pair source scheme etc. can be used here \cite{duan_long-distance_2001,barrett_efficient_2005,simon_quantum_2007,Sangouard}. Such a scheme would not need to use a QND detector. Classical communication of detection results to end stations, like in case of QND detection, is needed though. There can be many other variants of the repeater scheme which will have their individual benefits and challenges. 

Schemes employing both ground-based source and detectors has the same plethora of advantages described before. Among these the ability to create new experiments or make changes easily can specifically be important for quantum internet purposes. There can however be other forms of transmissions possible like pair-source transmission from satellites over which repeater scheme can be employed, similar to \cite{boone_entanglement_2015}, which may boost the rate further at the cost of further design complexity and flexibility. 

Another strategy towards quantum internet (not shown in figure) would be using the entanglement distribution by qubit transmission protocol described in Section \ref{scheme}. This would not need repeater protocol. In entanglement distribution by qubit transmission one photon of an entangled pairs is stored in a memory while the other is sent by qubit transmission to destination ground station. This protocol already needs quantum memory. By using another quantum memory in the destination ground station, along with a QND detector, will enable quantum Internet capabilities. This protocol would have both source and detector on ground and only one uplink is encountered like qubit transmission and hence there is only one turbulence loss. These features make this protocol another interesting candidate for implementing quantum internet.

Memories may be altogether not required in a certain case, with limited benefits, described in Fig. \ref{QI_Appendix_Fig}(c). A way of creating heralded entangled pairs would be send qubits redundantly while implementing quantum internet protocols to mitigate the effect of loss. That would have many challenges though. However, such redundancy approach would be important for small losses - say for only a few dB loss, e.g. for downlink where turbulence loss can be very small. Such small losses are possible, at least in principle, in our protocol, especially in entanglement source onboard satellite protocols (S1 in Fig. \ref{fig_scheme}(b)) containing two downlinks. Especially for small distances (below 5000 km), downlink entanglement distribution loss can be only a few dB. For very small losses the redundancy approach would essentially be equivalent to heralding approach as even while heralding, quantum memory and the subsequent quantum operations all would have some inherent losses themselves. The other reason to use memories - i.e. causality constraining us to wait for classical communication - can also be circumvented using the following technique. To perform a certain operation with an entangled photon pair, after a photonic qubit is measured at one end, the generated classical communication signal needs to reach the other end before the other qubit (the other photon of the entangled pair) reaches there. This can be achieved in entanglement distribution using a satellite source (S1 in Fig. \ref{fig_scheme}(b)) if the two photons emitted from the satellite travel unequal paths with difference between the two optical paths being greater than the classical communication optical path length. This is possible in 2D constellation of quantum satellites where one photon takes intentionally longer path through the 2D mesh, as seen in Fig. \ref{fig:quantum internet}(C). It needs to be remembered here that classical communication through optical fiber would be slower than satellite communication as light moves slower through glass. But as classical satellite internet with low latency LEO satellites is being available \cite{sheetz_spacex_2021}, that can be used. Using this scheme for example, an unknown qubit can be teleported it to a far away place, certain quantum operations can be done on it and the output qubit can be teleported back. The whole process would not require memory or QND detectors, although some redundancy would be required. The above proposal (not requiring memories) is complex though especially when a set of multiple qubits in unknown states (say, in an entangled state) is involved. Even for small amount of loss at each qubit the redundancy required increases very quickly, as the dependence is exponential. 

\subsection{Analysing different errors influencing diffraction}\label{error_appen}

\begin{figure}
     \centering
     \begin{subfigure}%[b]{0.3\textwidth}
         \centering
         \includegraphics[width=0.85\linewidth]{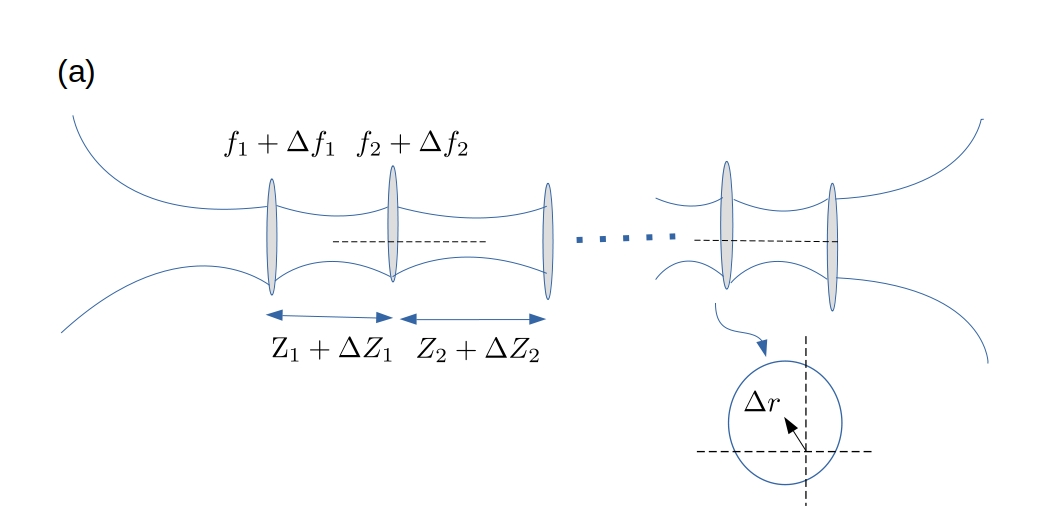} 
         %\caption{$y=x$}
         \label{fig:direc transm}
     \end{subfigure}
     \begin{subfigure}%[b]{0.3\textwidth}
         \centering
         \includegraphics[width=0.95\linewidth]{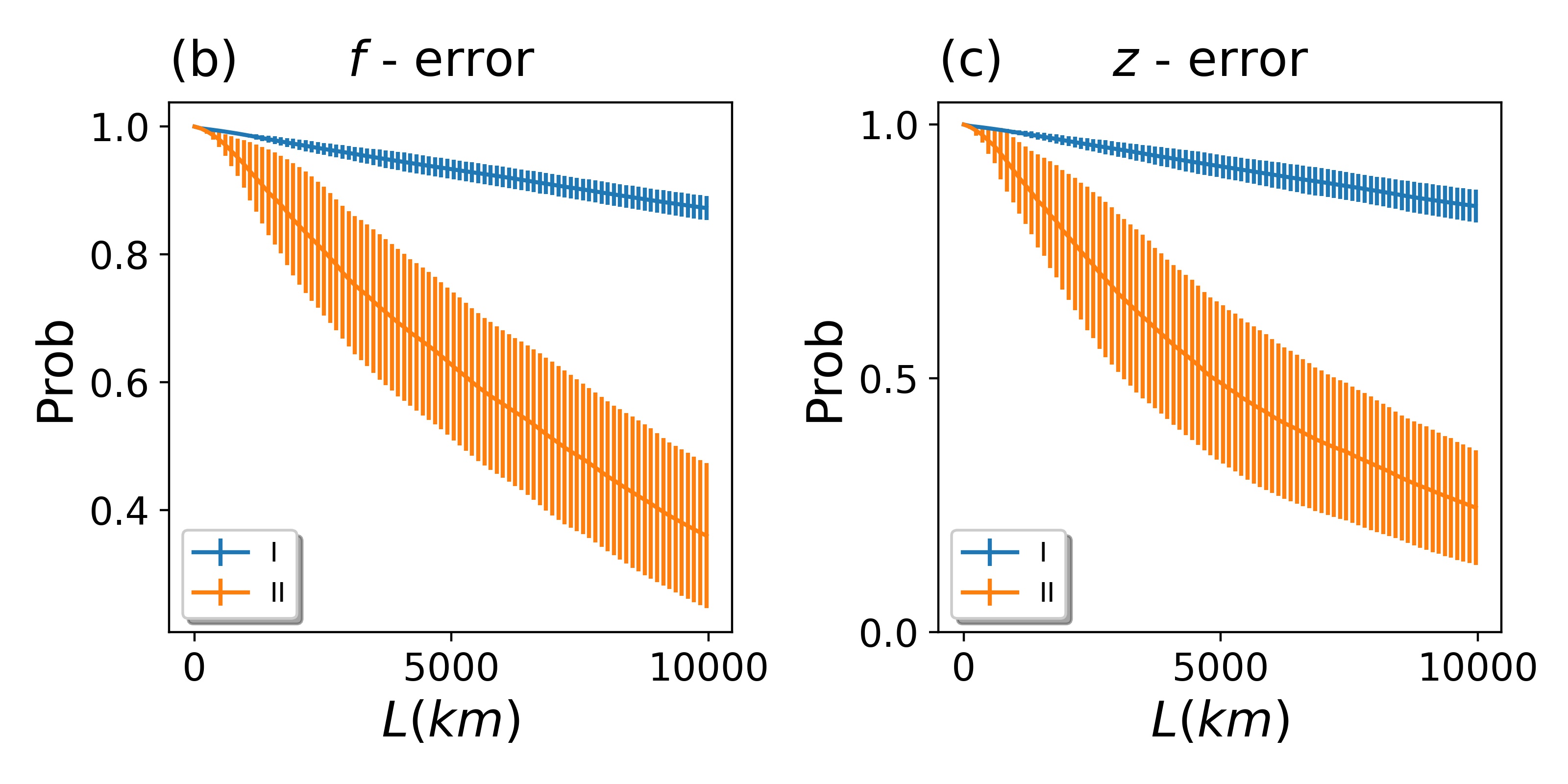}
         %\caption{$y=3sinx$}
         \label{fig:entangl_distr}
     \end{subfigure}
     \begin{subfigure}%[b]{0.3\textwidth}
         \centering
         \includegraphics[width=0.95\linewidth]{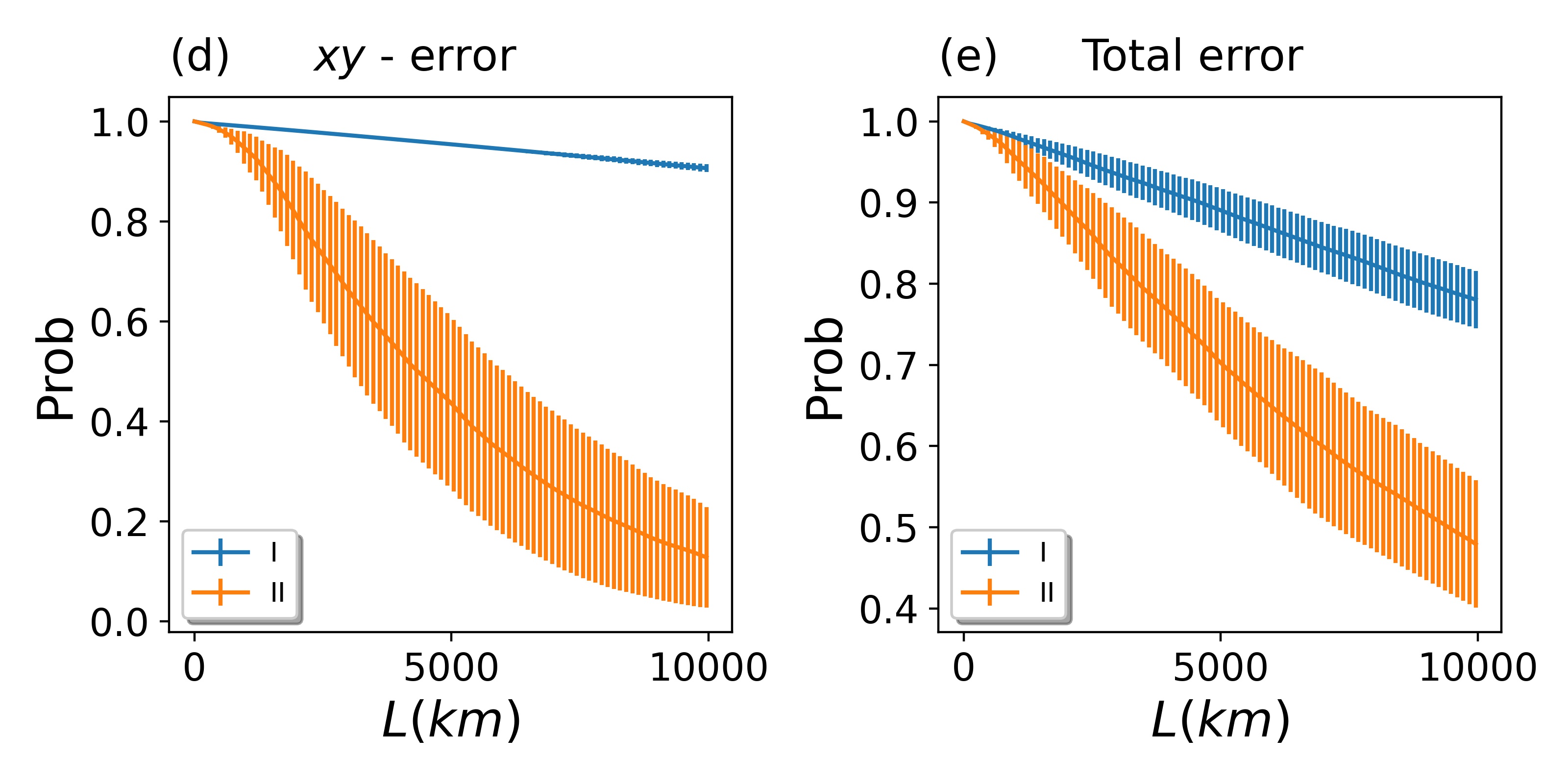}
         %\caption{$y=3sinx$}
         \label{fig:entangl_distr}
     \end{subfigure}
     \caption{ The effect of 'satellite lens' focal length ($f$) and position ($z$ or $xy$) error on beam propagation, (i.e., on diffraction loss) is shown. (a) Occurrence of the error processes are shown schematically.  (b-e) The effect of \textit{f} error (b), \textit{z} error (c) or \textit{xy} error (d) are shown each for two different amounts of error. In each case, transmission probability through 10,000 km is calculated over 300 repetitions using a uniform error distribution from where mean and standard deviation (std) is derived. The error values taken are 5$\%$ and 10$\%$ of $f$ for \textit{f} error, 5$\%$ and 20$\%$ of $L_0$ for \textit{z} error and 1$\%$ and 10$\%$ of $r$ ($d$/2) for \textit{xy} error. Combined effect of all these errors is shown in (e) for two different sets of parameter values. $f$,$z$ and $xy$ error values of (2.5$\%$, 5$\%$ and 1$\%$) results in transmission probability of 78$\%$ at 10k km while large errors of (5$\%$, 10$\%$ and 2$\%$) results in 48$\%$ transmission. For all plots $d$ = 60 cm, $L_0$ = 120 km and $\lambda$ = 800 nm.
}
    \label{Error_Appendix_Fig}
\end{figure}    

We discuss the effect of different satellite chain setup errors affecting diffraction below. We consider the effects of lens focal length error (\textit{f} error) end lens position errors in both $z$ direction (\textit{z} error) and $xy$ direction (\textit{xy} error). The different errors are shown schematically in Fig. \ref{Error_Appendix_Fig}(a). For example, focal length ($f$) error is modeled by introducing random errors in focal lengths ($\Delta f$), while $z$ or $xy$ errors are modeled by introducing a random longitudinal ($\Delta z$) or lateral displacement ($\Delta r$, $\Delta \theta$) error. Errors values are generated from an uniform random distributions of a given interval (e.g. 5$\%$ $f$ error means choosing random $f$ error values from the [-0.05$f$, 0.05$f$] interval). We have used 100 different error values in each case and calculated photon transmission probability at each satellite link for a total distance of 10,000 km. Hence, we had 100 different photon transmission probabilities for each distance. Their average and variance were calculated and we plotted the average photon transmission probability with distance while also showing the standard deviation at each point in each plot. The standard deviation values increased as the deteriorating effect of the errors increased (i.e. mean probability decreased) either due to large error values or with increasing distance or both. This can be easily understood. If there is no effect of errors (e.g. for negligibly small error values) there would be identical photon transmission value for all error values and thus variance would become zero.

In the following, we would primarily use the average photon transmission probability to assess the effect of the error. The effect of each of these errors on diffraction loss in entanglement distribution are considered first individually in Fig. \ref{Error_Appendix_Fig}(b)-(d) and then their combined effect is seen in Fig. \ref{Error_Appendix_Fig}(e). In all cases we considered $d$ = 60 cm $L_0$ = 120 km, and a total distance of $L$ = 10,000 km which is half of the global distance (20,000 km) and hence the distance one photon of an entangled pair needs to travel. Diffraction loss without any errors for the above parameters can be deduced to be 0.925, using Fig. \ref{Result_Fig1}(b). We first consider the effect of \textit{f} error. In Fig. \ref{Error_Appendix_Fig}(b) two cases are considered with uniform random focal length errors. These errors resided within intervals of $\pm 2.5 \%$ and $\pm 10\%$ of focal length ($f$), respectively. In each case, we simulated diffraction for 100 different runs of focal length (\textit{f}) errors. We calculated the mean and standard deviation of photon transmission probability at each aperture from these 100 runs which are plotted in Fig. \ref{Error_Appendix_Fig}(b). We see that for $2.5 \%$ \textit{f} error probability mean drops to $0.87$ while for $10\%$ error it drops to $0.36$. These probabilities are respectively 94 $\%$  and 39 $\%$  of the original photon transmission probability (0.925) without any errors. It shows 2.5 $\%$ \textit{f} error is quite acceptable and even a 10 $\%$ error doesn't have an extreme effect. The focal length error originates due to uncontrollable changes in focal length which can come either due to errors in adjusting mirror position for adjustable focal lengths setups or while manufacturing for fixed focal length setups. We discussed adjustable focal length in detail in Section \ref{telescope_setups} and derived the focal length formulas for different telescope setups of Fig. \ref{Diss_Fig1}(a)-(c). To achieve a focal length error below 10 $\%$ one needs a mirror position movement accuracy of 2 $\mu$ m for front mirrors in Fig. \ref{Diss_Fig1}(a) and Fig. \ref{Diss_Fig1}(c) while only 1 mm for Fig. \ref{Diss_Fig1}(b) although in absence of any front mirror the larger telescope (so  called) back mirrors need to be moved.

The error in satellite $z$ position is calculated similarly. Fig. \ref{Error_Appendix_Fig}(c) shows that in presence of $ 5\%$ \textit{z} error the mean photon transmission probability drops to $0.83$ while with $20\%$ \textit{z} error it drops to $ 0.24$. These probabilities are respectively 90 $\%$  and 26 $\%$  of the original photon transmission probability. A 20$\%$ \textit{z} error corresponds to 24 km \textit{z} error which is a very large error in satellite position. Satellite position error of 24 km in LEO orbit is a huge error which can be compensated and eventual irreducible  error would be much smaller, possibly even considerably smaller than 6 km, i.e. 5$\%$ error. 

The above simulations show that diffraction loss is rather robust against \textit{f} errors and \textit{z} errors which points to the fact that the set of focal length used are not unique and even with quite a bit of errors they can still confine the light well. This possibility to use a diverse number of lens setups to confine light was discussed earlier in context to lens wave guides in Section \ref{result_simulation} \cite{siegman_modes_1967, marcuse_propagation_1964}. \textit{f} error and \textit{z} error are also similar in the sense that a smaller focal length or a larger propagation distance influences the diffraction loss similarly. Hence, we can expect similar results for \textit{f} error and \textit{z} error.

Fig. \ref{Error_Appendix_Fig}(d) shows the error due to shift in lens centroid in \textit{xy} directions, i.e. \textit{xy} error. This can also be termed as $r-\theta$ error as the lens centroid error is calculated by taking random values in r (within a bound) and $\theta$. The two error values in Fig. \ref{Error_Appendix_Fig}(d) are given by $\Delta r = 0.3$ cm and $3$ cm which are $1\%$ and $10\%$ respectively of telescope radius R = 30 cm. \textit{xy} error is the largest of the three errors described here with photon transmission probability decreasing to $0.90 $ and $0.12$ at 10,000 kilometers which are 97 $\%$ and 13 $\%$ of the original case with no errors. Such a large error for 3 cm $xy$ error is understandable as a shifted lens not only truncates light more on one side but also deflects the beam at an angle. Such angular deflection causes the beam to go out of path and hence it gets even more truncated by subsequent lenses. One way such \textit{xy} position error can occur for the effective ‘satellite lens’ is by satellite position error. Evidently, even a moderate \textit{xy} error (in meters) due to position error would be completely devastating for the scheme. However, satellite positions errors in $xy$ can be easily eliminated by remembering the ‘satellite lenses’ are really made of mirrors. These mirrors can reflect light at a slightly different angle compensating for the $xy$ deviation. In fact, light is always reflected at an angle to account for the curvature of the earth as discussed earlier (Fig. \ref{Diss_Fig1}). One simply needs to adjust this angle slightly. This would result in slightly longer propagation distance and cause \textit{z} error. However, compensating this small $xy$ deviation (say, 1 m) in a large link (say, $z$ = 100 km) would cause completely negligible z error ($\sqrt{(10^5)^2+1}-10^5 \approx 5*10^{-6} m = 5 \micro m$). More importantly satellite $xy$ position error would actually produce an error in satellite velocity as it would change the orbital radius of the satellite. This would again produce \textit{z} errors and cause some difficulty in satellite alignment and tracking. However, orbital velocity (say, $v$) changes slowly with orbital radius (say, $R$), as $v \propto \frac{1}{R}$. Hence, 1 m change in orbital radius would result in only a 1 mm/s change in the original orbital velocity ($\sim$ 8 km/s in LEO).

Another important source of error related to $xy$ error is beam deviation error which would originate due to errors in telescope mirror alignment in the satellite chain. Slight changes in the transmitting telescope angle would deviate the light beam from the center of the receiving telescope and which would then further deviate the light beam. Beam deviation is an important sources of error. However, beam deviation error can't be faithfully simulated using our system of lenses as it originates from the mirrors used in the telescopes which can deflect light. For example, light can always be deflected by rotating a mirror, but a lens cannot deflect light if it falls at the centre of the lens. Only a shifted lens can deflect light. This implies not all aspects of beam deviation error can be modeled using a lens system with \textit{xy} error. However, \textit{xy} error is still similar to beam deviation error in many ways as shifted lenses do deflect light. Beam deviations also causes beam truncation as beams fall on telescope mirrors of centre. This effect is modelled properly in \textit{xy} error. 

Beam deviation error is one of the most important concerns in the existing quantum satellite experiments too where it manifests as tracking error. Tracking error for the Micius satellite was 0.41 $\mu$ rad \cite{yin_satellite-based_2017}. Such an error would result in a $xy$ error of 4.1 cm for 100 km satellite separation. Considering 0.41 $\mu$ rad was dynamic tracking error in a satellite-ground link, alignment error in a relatively stationary satellite chain should be considerably less. Hence, the $xy$ error values used in the simulation above (0.3 cm and 3 cm) were within reasonable limits. Hence, \textit{xy} error simulation do gives us some ideas about the effect of beam deviation error.

In Fig. \ref{Error_Appendix_Fig}(e) the combined effect of $f$, $z$ and $xy$ errors in photon transmission is investigated. In one case, $f$, $z$ and $xy$ errors of 2.5$\%$, 5$\%$, 1$\%$ respectively generated an average photon transmission probability of 0.78, while in another case 5$\%$, 10$\%$ and 2$\%$ respective errors generated a mean probability of 0.48. These mean probabilities of photon transmission (at 10,000 km), in presence of errors, are respectively 84 $\%$ ( 0.76 dB) and 51 $\%$ ( 2.92 dB) of the 92.5 $\%$ photon transmission probability, while not considering any error. Thus, the different satellite change setup errors do affect photon transmission probability, although even their combined effect can probably be constrained to only a few dB of extra loss if the above-described system parameter regimes can be achieved.

% \bibliographystyle{unsrt}
% \bibliography{sat_QN_ref}

\end{document}